\documentclass[twocolumn]{aastex63}

\usepackage{mathtools}
\usepackage{CJK}

\newcommand{\ra}[4]{$#1\overset{\mathrm h}{\phd}#2\overset{\mathrm m}{\phd}#3\overset{\mathrm s}{.}#4$}
\newcommand{\dec}[4]{$#1\overset{\circ}{\phd}#2\overset{\prime}{\phd}#3\overset{\prime\prime}{.}#4$}

\usepackage[multiple]{footmisc}
\usepackage{soul}
\newcommand\N[1]{{\color{blue} \bf #1}} 

\received{}
\revised{}
\accepted{}

\shorttitle{}
\shortauthors{}

\graphicspath{{./}{figures/}}

\begin{document}

\title{Circumstellar Medium Interaction in SN~2018lab, A Low-Luminosity II-P Supernova observed with TESS}

\correspondingauthor{Jeniveve Pearson}
\email{jenivevepearson@arizona.edu}

\newcommand{\LCO}{\affiliation{Las Cumbres Observatory, 6740 Cortona Drive, Suite 102, Goleta, CA 93117-5575, USA}}
\newcommand{\UCSB}{\affiliation{Department of Physics, University of California, Santa Barbara, CA 93106-9530, USA}}
\newcommand{\KITP}{\affiliation{Kavli Institute for Theoretical Physics, University of California, Santa Barbara, CA 93106-4030, USA}}
\newcommand{\UCD}{\affiliation{Department of Physics and Astronomy, University of California, Davis, 1 Shields Avenue, Davis, CA 95616-5270, USA}}
\newcommand{\WIS}{\affiliation{Department of Particle Physics and Astrophysics, Weizmann Institute of Science, 76100 Rehovot, Israel}}
\newcommand{\OKC}{\affiliation{Oskar Klein Centre, Department of Astronomy, Stockholm University, Albanova University Centre, SE-106 91 Stockholm, Sweden}}
\newcommand{\OAPD}{\affiliation{INAF-Osservatorio Astronomico di Padova, Vicolo dell'Osservatorio 5, I-35122 Padova, Italy}}
\newcommand{\Caltech}{\affiliation{Cahill Center for Astronomy and Astrophysics, California Institute of Technology, Mail Code 249-17, Pasadena, CA 91125, USA}}
\newcommand{\GSFC}{\affiliation{Astrophysics Science Division, NASA Goddard Space Flight Center, Mail Code 661, Greenbelt, MD 20771, USA}}
\newcommand{\UMD}{\affiliation{Joint Space-Science Institute, University of Maryland, College Park, MD 20742, USA}}
\newcommand{\UCB}{\affiliation{Department of Astronomy, University of California, Berkeley, CA 94720-3411, USA}}
\newcommand{\TTU}{\affiliation{Department of Physics, Texas Tech University, Box 41051, Lubbock, TX 79409-1051, USA}}
\newcommand{\STScI}{\affiliation{Space Telescope Science Institute, 3700 San Martin Drive, Baltimore, MD 21218-2410, USA}}
\newcommand{\UT}{\affiliation{University of Texas at Austin, 1 University Station C1400, Austin, TX 78712-0259, USA}}
\newcommand{\IoA}{\affiliation{Institute of Astronomy, University of Cambridge, Madingley Road, Cambridge CB3 0HA, UK}}
\newcommand{\QUB}{\affiliation{Astrophysics Research Centre, School of Mathematics and Physics, Queen's University Belfast, Belfast BT7 1NN, UK}}
\newcommand{\IPAC}{\affiliation{Spitzer Science Center, California Institute of Technology, Pasadena, CA 91125, USA}}
\newcommand{\JPL}{\affiliation{Jet Propulsion Laboratory, California Institute of Technology, 4800 Oak Grove Dr, Pasadena, CA 91109, USA}}
\newcommand{\Southampton}{\affiliation{Department of Physics and Astronomy, University of Southampton, Southampton SO17 1BJ, UK}}
\newcommand{\LANL}{\affiliation{Space and Remote Sensing, MS B244, Los Alamos National Laboratory, Los Alamos, NM 87545, USA}}
\newcommand{\Tsinghua}{\affiliation{Physics Department and Tsinghua Center for Astrophysics, Tsinghua University, Beijing, 100084, People's Republic of China}}
\newcommand{\NAOC}{\affiliation{National Astronomical Observatory of China, Chinese Academy of Sciences, Beijing, 100012, People's Republic of China}}
\newcommand{\Itagaki}{\affiliation{Itagaki Astronomical Observatory, Yamagata 990-2492, Japan}}
\newcommand{\Einstein}{\altaffiliation{Einstein Fellow}}
\newcommand{\Hubble}{\altaffiliation{Hubble Fellow}}
\newcommand{\CfA}{\affiliation{Center for Astrophysics \textbar{} Harvard \& Smithsonian, 60 Garden Street, Cambridge, MA 02138-1516, USA}}
\newcommand{\UA}{\affiliation{Steward Observatory, University of Arizona, 933 North Cherry Avenue, Tucson, AZ 85721-0065, USA}}
\newcommand{\MPIA}{\affiliation{Max-Planck-Institut f\"ur Astrophysik, Karl-Schwarzschild-Stra\ss{}e 1, D-85748 Garching, Germany}}
\newcommand{\DSFP}{\altaffiliation{LSSTC Data Science Fellow}}
\newcommand{\HCO}{\affiliation{Harvard College Observatory, 60 Garden Street, Cambridge, MA 02138-1516, USA}}
\newcommand{\Carnegie}{\affiliation{Observatories of the Carnegie Institute for Science, 813 Santa Barbara Street, Pasadena, CA 91101-1232, USA}}
\newcommand{\TAU}{\affiliation{School of Physics and Astronomy, Tel Aviv University, Tel Aviv 69978, Israel}}
\newcommand{\Edinburgh}{\affiliation{Institute for Astronomy, University of Edinburgh, Royal Observatory, Blackford Hill EH9 3HJ, UK}}
\newcommand{\Birmingham}{\affiliation{Birmingham Institute for Gravitational Wave Astronomy and School of Physics and Astronomy, University of Birmingham, Birmingham B15 2TT, UK}}
\newcommand{\Bath}{\affiliation{Department of Physics, University of Bath, Claverton Down, Bath BA2 7AY, UK}}
\newcommand{\CTIO}{\affiliation{Cerro Tololo Inter-American Observatory, National Optical Astronomy Observatory, Casilla 603, La Serena, Chile}}
\newcommand{\Potsdam}{\affiliation{Institut f\"ur Physik und Astronomie, Universit\"at Potsdam, Haus 28, Karl-Liebknecht-Str. 24/25, D-14476 Potsdam-Golm, Germany}}
\newcommand{\INPE}{\affiliation{Instituto Nacional de Pesquisas Espaciais, Avenida dos Astronautas 1758, 12227-010, S\~ao Jos\'e dos Campos -- SP, Brazil}}
\newcommand{\UNC}{\affiliation{Department of Physics and Astronomy, University of North Carolina, 120 East Cameron Avenue, Chapel Hill, NC 27599, USA}}
\newcommand{\Ohio}{\affiliation{Astrophysical Institute, Department of Physics and Astronomy, 251B Clippinger Lab, Ohio University, Athens, OH 45701-2942, USA}}
\newcommand{\AAS}{\affiliation{American Astronomical Society, 1667 K~Street NW, Suite 800, Washington, DC 20006-1681, USA}}
\newcommand{\MMT}{\affiliation{MMT and Steward Observatories, University of Arizona, 933 North Cherry Avenue, Tucson, AZ 85721-0065, USA}}
\newcommand{\Geneva}{\affiliation{ISDC, Department of Astronomy, University of Geneva, Chemin d'\'Ecogia, 16 CH-1290 Versoix, Switzerland}}
\newcommand{\IUCAA}{\affiliation{Inter-University Center for Astronomy and Astrophysics, Post Bag 4, Ganeshkhind, Pune, Maharashtra 411007, India}}
\newcommand{\CMU}{\affiliation{Department of Physics, Carnegie Mellon University, 5000 Forbes Avenue, Pittsburgh, PA 15213-3815, USA}}
\newcommand{\NAOJ}{\affiliation{Division of Science, National Astronomical Observatory of Japan, 2-21-1 Osawa, Mitaka, Tokyo 181-8588, Japan}}
\newcommand{\IfA}{\affiliation{Institute for Astronomy, University of Hawai`i, 2680 Woodlawn Drive, Honolulu, HI 96822-1839, USA}}
\newcommand{\UCSC}{\affiliation{Department of Astronomy and Astrophysics, University of California, Santa Cruz, CA 95064-1077, USA}}
\newcommand{\Purdue}{\affiliation{Department of Physics and Astronomy, Purdue University, 525 Northwestern Avenue, West Lafayette, IN 47907-2036, USA}}
\newcommand{\Princeton}{\affiliation{Department of Astrophysical Sciences, Princeton University, 4 Ivy Lane, Princeton, NJ 08540-7219, USA}}
\newcommand{\Moore}{\affiliation{Gordon and Betty Moore Foundation, 1661 Page Mill Road, Palo Alto, CA 94304-1209, USA}}
\newcommand{\Durham}{\affiliation{Department of Physics, Durham University, South Road, Durham, DH1 3LE, UK}}
\newcommand{\JHU}{\affiliation{Department of Physics and Astronomy, The Johns Hopkins University, 3400 North Charles Street, Baltimore, MD 21218, USA}}
\newcommand{\Toronto}{\affiliation{David A.\ Dunlap Department of Astronomy and Astrophysics, University of Toronto,\\ 50 St.\ George Street, Toronto, Ontario, M5S 3H4 Canada}}
\newcommand{\Duke}{\affiliation{Department of Physics, Duke University, Campus Box 90305, Durham, NC 27708, USA}}
\newcommand{\NCU}{\affiliation{Graduate Institute of Astronomy, National Central University, 300 Jhongda Road, 32001 Jhongli, Taiwan}}
\newcommand{\Columbia}{\affiliation{Department of Physics and Columbia Astrophysics Laboratory, Columbia University, Pupin Hall, New York, NY 10027, USA}}
\newcommand{\Flatiron}{\affiliation{Center for Computational Astrophysics, Flatiron Institute, 162 5th Avenue, New York, NY 10010-5902, USA}}
\newcommand{\CIERA}{\affiliation{Center for Interdisciplinary Exploration and Research in Astrophysics and Department of Physics and Astronomy, \\Northwestern University, 1800 Sherman Avenue, 8th Floor, Evanston, IL 60201, USA}}
\newcommand{\GeminiNorth}{\affiliation{Gemini Observatory, 670 North A`ohoku Place, Hilo, HI 96720-2700, USA}}
\newcommand{\Keck}{\affiliation{W.~M.~Keck Observatory, 65-1120 M\=amalahoa Highway, Kamuela, HI 96743-8431, USA}}
\newcommand{\UW}{\affiliation{Department of Astronomy, University of Washington, 3910 15th Avenue NE, Seattle, WA 98195-0002, USA}}
\newcommand{\DiRAC}{\altaffiliation{DiRAC Fellow}}
\newcommand{\USask}{\affiliation{Department of Physics \& Engineering Physics, University of Saskatchewan, 116 Science Place, Saskatoon, SK S7N 5E2, Canada}}
\newcommand{\Thacher}{\affiliation{Thacher School, 5025 Thacher Road, Ojai, CA 93023-8304, USA}}
\newcommand{\Rutgers}{\affiliation{Department of Physics and Astronomy, Rutgers, the State University of New Jersey,\\136 Frelinghuysen Road, Piscataway, NJ 08854-8019, USA}}
\newcommand{\FSU}{\affiliation{Department of Physics, Florida State University, 77 Chieftan Way, Tallahassee, FL 32306-4350, USA}}
\newcommand{\Melbourne}{\affiliation{School of Physics, The University of Melbourne, Parkville, VIC 3010, Australia}}
\newcommand{\ASTROthreeD}{\affiliation{ARC Centre of Excellence for All Sky Astrophysics in 3 Dimensions (ASTRO 3D)}}
\newcommand{\Stromlo}{\affiliation{Mt.\ Stromlo Observatory, The Research School of Astronomy and Astrophysics, Australian National University, ACT 2601, Australia}}
\newcommand{\NCPAS}{\affiliation{National Centre for the Public Awareness of Science, Australian National University, Canberra, ACT 2611, Australia}}
\newcommand{\TAMU}{\affiliation{Department of Physics and Astronomy, Texas A\&M University, 4242 TAMU, College Station, TX 77843, USA}}
\newcommand{\Mitchell}{\affiliation{George P.\ and Cynthia Woods Mitchell Institute for Fundamental Physics \& Astronomy, College Station, TX 77843, USA}}
\newcommand{\ESO}{\affiliation{European Southern Observatory, Alonso de C\'ordova 3107, Casilla 19, Santiago, Chile}}
\newcommand{\ICE}{\affiliation{Institute of Space Sciences (ICE, CSIC), Campus UAB, Carrer
de Can Magrans, s/n, E-08193 Barcelona, Spain}}
\newcommand{\IEEC}{\affiliation{Institut d'Estudis Espacials de Catalunya, Gran Capit\`a, 2-4, Edifici Nexus, Desp.\ 201, E-08034 Barcelona, Spain}}
\newcommand{\Warwick}{\affiliation{Department of Physics, University of Warwick, Gibbet Hill Road, Coventry CV4 7AL, UK}}
\newcommand{\Macquarie}{\affiliation{School of Mathematical and Physical Sciences, Macquarie University, NSW 2109, Australia}}
\newcommand{\AAARC}{\affiliation{Astronomy, Astrophysics and Astrophotonics Research Centre, Macquarie University, Sydney, NSW 2109, Australia}}
\newcommand{\Capodimonte}{\affiliation{INAF - Capodimonte Astronomical Observatory, Salita Moiariello 16, I-80131 Napoli, Italy}}
\newcommand{\INFNNapoli}{\affiliation{INFN - Napoli, Strada Comunale Cinthia, I-80126 Napoli, Italy}}
\newcommand{\ICRANet}{\affiliation{ICRANet, Piazza della Repubblica 10, I-65122 Pescara, Italy}}
\newcommand{\MSU}{\affiliation{Center for Data Intensive and Time Domain Astronomy, Department of Physics and Astronomy,\\Michigan State University, East Lansing, MI 48824, USA}}
\newcommand{\SETI}{\affiliation{SETI Institute,
339 Bernardo Ave, Suite 200, Mountain View, CA 94043, USA}}
\newcommand{\IAIFI}{\affiliation{The NSF AI Institute for Artificial Intelligence and Fundamental Interactions}}
\author[0000-0002-0744-0047]{Jeniveve Pearson}
\UA
\author[0000-0002-0832-2974]{Griffin Hosseinzadeh}
\UA
\author[0000-0003-4102-380X]{David J. Sand}
\UA
\author[0000-0003-0123-0062]{Jennifer E. Andrews}
\GeminiNorth
\author[0000-0001-5754-4007]{Jacob E. Jencson}
\UA
\author[0000-0002-7937-6371]{Yize Dong \begin{CJK*}{UTF8}{gbsn}(董一泽)\end{CJK*}}
\UCD
\author[0000-0002-4924-444X]{K. Azalee Bostroem}
\UW
\author[0000-0001-8818-0795]{S.~Valenti}
\UCD
\author[0000-0003-0549-3281]{Daryl Janzen}
\USask
\author[0000-0002-7015-3446]{Nicol\'as Meza Retamal}
\UCD
\author[0000-0001-9589-3793]{M.~J. Lundquist}
\Keck
\author[0000-0003-2732-4956]{Samuel Wyatt}
\UA
\author[0000-0002-1546-9763]{R.~C. Amaro}
\UA
\author[0000-0003-0035-6659]{Jamison Burke}
\LCO\UCSB
\author[0000-0003-4253-656X]{D.\ Andrew Howell}
\LCO\UCSB
\author[0000-0001-5807-7893]{Curtis McCully}
\LCO\UCSB
\author[0000-0002-1125-9187]{Daichi Hiramatsu}
\CfA\IAIFI
\author[0000-0001-8738-6011]{Saurabh W.\ Jha}
\Rutgers
\author[0000-0001-5510-2424]{Nathan Smith}
\UA
\author[0000-0002-6703-805X]{Joshua Haislip}
\UNC
\author[0000-0003-3642-5484]{Vladimir Kouprianov}
\UNC
\author[0000-0002-5060-3673]{Daniel E.\ Reichart}
\UNC
\author[0000-0002-6535-8500]{Yi Yang}
\UCB
\author[0000-0003-3643-839X]{Jeonghee Rho}
\SETI

\begin{abstract}

We present photometric and spectroscopic data of SN~2018lab, a low luminosity type IIP supernova (LLSN) with a V-band peak luminosity of $-15.1\pm0.1$ mag. 
SN~2018lab was discovered by the Distance Less Than 40 Mpc (DLT40) SNe survey only 0.73 days post-explosion, as determined by observations from the Transiting Exoplanet Survey Satellite (TESS).
TESS observations of SN~2018lab yield a densely sampled, fast-rising, early time light curve likely powered by circumstellar medium (CSM) interaction. The blue-shifted, broadened flash feature in the earliest spectra ($<$2 days) of SN~2018lab provide further evidence for ejecta-CSM interaction. 
The early emission features in the spectra of SN~2018lab are well described by models of a red supergiant progenitor with an extended envelope and close-in CSM. As one of the few LLSNe with observed flash features, SN~2018lab highlights the need for more early spectra to explain the diversity of flash feature morphology in type II SNe.

\end{abstract}


\section{Introduction} \label{sec:intro}

Type IIP/IIL supernovae (SNe II) are the result of core-collapse in stars $>$8 M$_{\odot}$, and are defined by the appearance of hydrogen in their spectra \citep{Filippenko1997, Smartt2009}.
SNe II have proven to be a continuous population smoothly spanning a significant photometric, $-19.0\lesssim M_\mathrm{V}\lesssim-13.0$ mag at peak, and spectroscopic diversity \citep{Anderson2014, Sanders2015, Valenti2016, Gutierrez2017}. The extrema of the SNe II distribution have been the subject of intense study. 
SNe II with peak magnitudes $M_V\ge-15.5$ are referred to as Low Luminosity (LL) SNe \citep{Pastorello2004}. The plateau luminosities of SNe II correlate with their photospheric expansion velocities \citep{Hamuy2002, Pejcha2015}. In line with this relation, LLSNe have the lowest expansion speeds \citep[$\sim1300-2500$ km s$^{-1}$ at 50 days post-explosion,][]{Pastorello2004, Spiro2014} of all SNe II. LLSNe also have smaller ejecta kinetic energies \citep[$\sim0.1-0.5 \times 10^{51}$ ergs,][]{Pumo2017} and lower nickel masses \citep[$\le10^{-2}$ M$_\odot$,][]{Turatto1998, Pastorello2004, Spiro2014} than typical SNe II \citep{Pastorello2004}.

The progenitors of LLSNe are unclear, despite their similarities to more luminous SNe II. The controversy surrounding the progenitors of LLSNe began with the discovery and subsequent progenitor modeling of SN~1997D \citep{Turatto1998, Benetti2001}. The characteristics of SN~1997D were well-explained by models of both the core collapse of a $>$20 M$_\odot$ star with a large amount of fallback \citep{Turatto1998, Zampieri1998} and of a star near the mass limit for undergoing core-collapse \citep[8-10 M$_\odot$,][]{Chugai2000}. In the time since, studies have supported both high \citep[$>$20 M$_\odot$]{Zampieri2003} and low-mass \citep[8-10 M$_\odot$]{Pignata2013, Pumo2017, 08bk, Lisakov2018, Kozyreva2022} red supergiant (RSG) progenitor models. Models with less massive (8-10 M$_\odot$) progenitors have become popular in recent years as archival pre-explosion Hubble Space Telescope (HST) images have placed upper limits on the progenitor masses of numerous LLSNe \citep{VanDyk2003, VanDyk2012, Maund2005, Li2006, Smartt2009, Fraser2011, Maund2014}.

Electron-capture (EC) SNe, the result of O-Ne-Mg core collapse in super-Asymptotic Giant Branch (AGB) stars, have also been used to explain the properties of some LLSNe \citep{16bkv, 21aai}. Some models predict that ECSNe can appear nearly identical to low luminosity core collapse (CC) SNe \citep{Nomoto1984, Kitaura2006, Poelarends2008} and their progenitors lie in the same mass range \citep[super-AGB stars 8-10 M$_\odot$,][]{Kitaura2006} as low mass RSGs which undergo core-collapse. Reliably distinguishing between the ECSNe and low luminosity CCSNe populations remains a challenge \citep{Zhang2020, 18zdEC, Callis2021}. 

All massive stars are expected to lose mass, however the properties of mass loss (e.g. density, radial extent, physical location) vary for different progenitors  \citep{smith14}.
Therefore, the extent of ejecta-CSM interaction is a possible indicator of whether an LLSN is from an electon-capture or core-collapse. Super-AGB stars readily produce large CSM envelopes as a result of their thermal pulsation phase. RSGs often have nearby CSM as well, though often much less than super-AGB stars, due to late stage episodic and eruptive mass loss. Indicators of ejecta-CSM interaction are sometimes only visible in the hours and days immediately following a SN explosion, before the SN ejecta has completely overtaken any CSM. Ejecta-CSM interaction can result in increased luminosity within the first weeks following explosion, observed as a bump or fast rise in the early light curve \citep{Anderson2014, GonzalezGaitan2015, Valenti2016, Morozova2017, Morozova2018, Forster2018, Hiramatsu2021}. More dense and substantial CSM will result in a larger -- and possibly longer -- excess luminosity and a greater effect on the early light curve. 
LLSNe with pronounced early time light curve bumps, like SN~2016bkv \citep{16bkv}, may have super-AGB progenitors.

Narrow emission lines observed in the spectra of SNe in the days following explosion can be used to indicate the composition, density, and velocity of the CSM surrounding the progenitor \citep{GalYam2014}. These narrow lines, often referred to as ``flash'' spectroscopy, are the result of recombination of CSM ionized by the shock-breakout flash \citep{Khazov2016} or very early ejecta-CSM interaction \citep{Smith2015, Shivvers2015} which ends when the CSM is entirely swept up by the expanding ejecta. 

Narrow lines from ionized CSM have been detected in the hours following explosion in some instances \citep{Niemela1985, Benetti1994, Quimby2007, GalYam2014}. When these spectral features are detected, they can provide insight into the composition and mass-loss history of the progenitor \citep{Groh2014, Yaron2017, Davies2019}. To date the only LLSN that clearly exhibits narrow early time flash features is SN~2016bkv \citep{16bkv}. 

A few SNe have shown signs of broadened, blue-shifted features rather than narrow ones in the days following explosion \citep{Soumagnac2020, Bruch2021, 21yja}. These broad features, hereafter referred to as broad-lined flash features, are produced when the outer layers of SN ejecta interact with low density CSM. 
The substantial CSM surrounding a super-AGB progenitor is likely to produce narrow lines at $\sim$5 days which can persist for up to a week, whereas CSM surrounding a RSG progenitor produces flash features which are typically expected to broaden and fade by $\sim$5 days post-explosion \citep{18zdEC}. 
We must emphasize that the prolonged existence of narrow-lined flash features in and of itself does not distinguish super-AGB progenitors from RSG progenitors. Some SNe with RSG progenitors exhibit narrow-lined flash features that remain visible for over a week (SN~1998S, \citealt{98S_1}, \citealt{98S_2}; SN~2020tlf, \citealt{20tlf}) and while the suggested ECSN SN~2018zd has long-lived narrow-lined flash features \citep{18zdEC}, SN~2018zd might be a CCSN with a RSG progenitor \citep{Zhang2020, Callis2021}.
However, no LLSN with narrow-lined flash features has a confirmed RSG progenitor. So in LLSNe, short-lived, early-time, broadened, blue-shifted flash features could be an indicator of a RSG progenitor rather than a super-AGB one, assuming there is no extreme long-term mass loss around the RSG. 

In this work, we present spectroscopic and photometric data for SN~2018lab, a LLSN which displays clear signs of CSM interaction: a fast rising light curve and a broad-lined flash feature in the early spectra ($<$2 days). 
In Section \ref{sec:disc} the discovery and classification of SN~2018lab is reviewed. In Section \ref{sec:obs} the observations and data reduction are outlined. In Section \ref{sec:lcprop} the photometric evolution is discussed. In Section \ref{sec:Spec} we present the spectroscopic evolution. These results are summarized in Section \ref{sec:summary}. 

\section{Discovery and Classification}\label{sec:disc}

\begin{figure}
    \centering
    \includegraphics[width=\hsize]{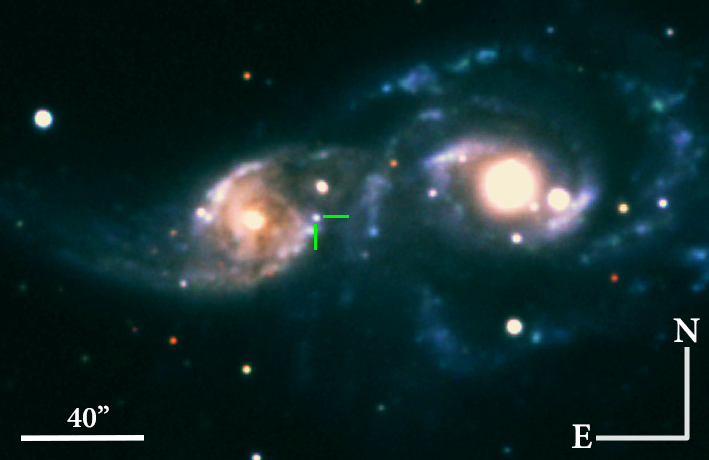}
    \caption{Composite $g,r,i$ image of SN~2018lab (indicated by green tick marks) in IC 2163 (left) obtained by Las Cumbres Observatory on 2019 January 10. NGC 2207 is on the right.}
    \label{fig:Loc}
\end{figure}

SN~2018lab, also known as DLT18ar, was first discovered, at RA(2000) = \ra{06}{16}{26}{520} and Dec(2000) = \dec{-21}{22}{32}{38}, by the Distance Less Than 40 Mpc Survey \citep[DLT40, for survey details see][]{Tartaglia2018} on 2018-12-29 at 03:01:26 UTC \citep[58481.126 MJD;][]{18labdis}. 

The redshift of SN~2018lab is $z_\mathrm{18lab} = 0.0089$, measured using the host H$\alpha$ in the first spectrum (1.6 days after explosion). SN~2018lab is located between the interacting galaxies IC 2163 and NGC 2207 (see Figure~\ref{fig:Loc}). IC 2163 and NGC 2207 are a well-studied pair of interacting, grazing galaxies \citep{Elmegreen1995a, Elmegreen1995b, Elmegreen2001, Elmegreen2006, Struck2005} that frequently produce SNe, notably SN~1975A \citep{75a, Arnett1982}, SN~2003H \citep{03H, 03H2}, SN~2010jp \citep{10jp, Corgan2022}, SN~2013ai \citep{13ai}, and SPIRITS 14buu, 15c and 17lb \citep{Jencson2017,jencson2019}.
IC 2163 has a redshift $z=0.0090$ \citep{2163z} and NGC 2207 has a redshift $z=0.0092$ \citep{2207z}. The measured redshift to SN~2018lab is most consistent with that of IC 2163, which is quoted as the host galaxy throughout this work.

IC 2163 was in the Transiting Exoplanet Survey Satellite \citep[TESS;][]{Ricker2015} footprint when SN~2018lab exploded. TESS observations of SN~2018lab yield an explosion date of MJD 58480.4$\pm$0.1, as published in \citet[][see their Eq. 2 and Table 1]{TESSdata}. This explosion time is 0.24 days after the last DLT40 non-detection and 0.73 days before DLT40's discovery of SN~2018lab, as seen in Figure \ref{fig:tess}.  We adopt the TESS-derived explosion epoch throughout this work. Spectroscopic classification done on 2018-12-31 at 06:42:29 UTC, 2 days after the explosion, confirmed that the object was an SN II \citep{18labclass}. 

\section{Observations and Data Reduction}\label{sec:obs}

\subsection{Follow-up Photometry and Spectroscopy}
\subsubsection{Photometry}
\begin{figure}
    \centering
    \includegraphics[width=\hsize]{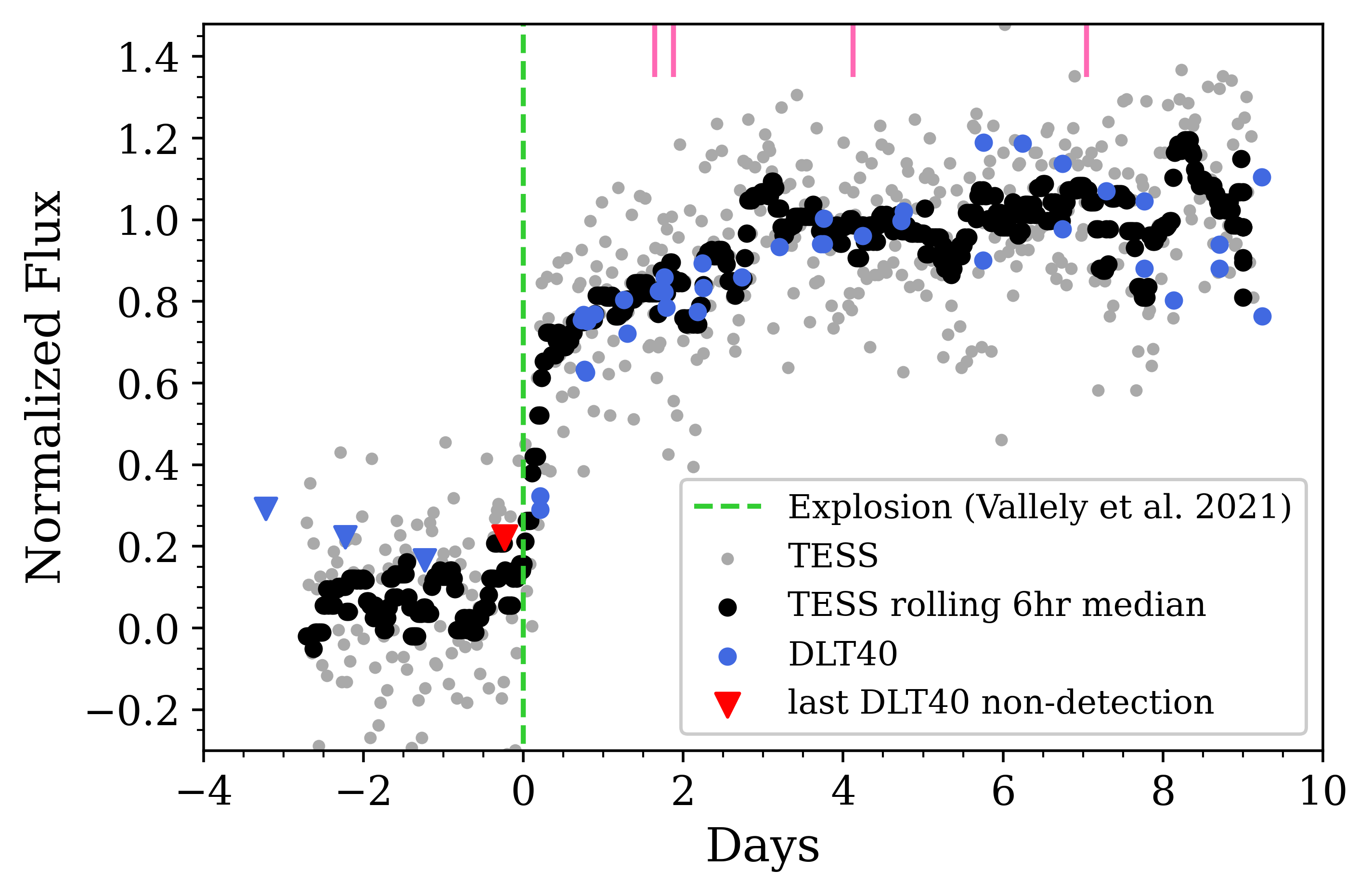}
    \caption{TESS-band SN~2018lab light curve (gray and black) and r-band corrected DLT40 (blue and red) data. Times of spectral epochs are marked in pink ticks at the top. The explosion time as reported in \citet{TESSdata} is marked by the green dashed line. This value is extremely well constrained by both TESS and DLT40 observations  (see Figure \ref{fig:lc_all} for zoomed in version). The light curves are normalized to the median value of the data 3--8 days after explosion.}
    \label{fig:tess}
\end{figure}

SN~2018lab was observed by TESS during the mission's Sector 6 operations, from 2018-12-15 18:36:03.542 to 2019-01-06 12:36:19.181 UTC. The TESS lightcurve of SN~2018lab was previously published in \citet{TESSdata}. In Figure \ref{fig:tess}, the TESS photometry, both unbinned and rolling 6-hr medians, is plotted.

Following the discovery of SN~2018lab by the DLT40 survey, continued monitoring was done by two of DLT40's discovery telescopes, the PROMPT5 0.4m telescope at the Cerro Tololo Inter-American Observatory and the PROMPT-MO 0.4m telescope at the Meckering Observatory in Australia. Observations taken by these telescopes are calibrated to the SDSS \textit{r} band, as described in \citet{Tartaglia2018}, and are shown in Figure \ref{fig:tess}. 

\begin{figure*}
    \centering
    \includegraphics[width=\hsize]{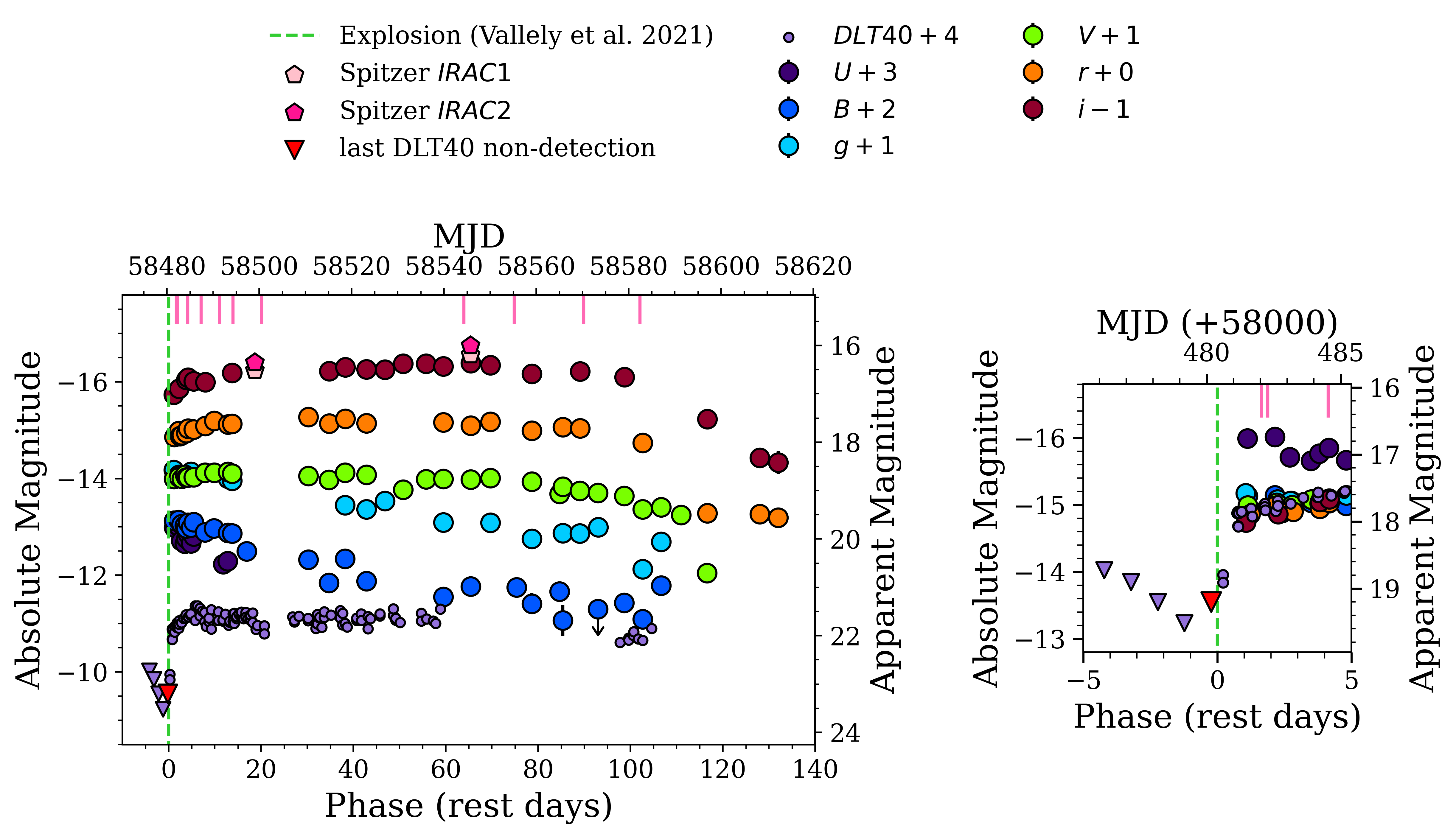}
    \caption{SN~2018lab light curves from DLT40, Las Cumbres Observatory, and Spitzer/IRAC. Left: light curves with offsets. Right: zoom in of the light curve in the first 5 days after explosion without offsets. Spectroscopic epochs are shown as pink lines along the upper x-axis.}
    \label{fig:lc_all}
\end{figure*}

Additional \textit{UBVgri} photometry of SN~2018lab was obtained using the Sinistro cameras on Las Cumbres Observatory's robotic 1m telescopes \citep{LasCumbres}, located at the Siding Spring Observatory, the South African Astronomical Observatory, and the Cerro Tololo Inter-American Observatory. These are shown in Figure \ref{fig:lc_all}

The photometric data from Las Cumbres Observatory was reduced using \texttt{lcogtsnpipe} \citep{Valenti2016}, a PyRAF-based image reduction pipeline.
Given the complexity of the host, \textit{UBVgri} reference images were obtained with Las Cumbres Observatory on 2021 August 25, $>$900 days after explosion, when the SN was no longer bright enough to be detectable. These reference frames were subtracted from the science images. Aperture photometry was then extracted from the difference images using \texttt{lcogtsnpipe}. Apparent magnitudes were calibrated to the APASS (\textit{BVgri}) catalog and Landolt (\textit{U}) standard fields observed on the same nights with the same telescopes.

Infrared photometry of SN~2018lab was also obtained with images from the Infrared Array Camera \citep[IRAC,][]{fazio2004} on board the Spitzer Space Telescope \citep{werner2004,gehrz2007}. The host system was imaged several times between 2014--2019 in the IRAC1 ($3.6$~$\mu$m) and IRAC2 ($4.5$~$\mu$m) imaging bands by the SPitzer InfraRed Intensive Transients Survey (SPIRITS; PI M.\ Kasliwal; PIDs 10136, 11063, 13053, 14089). The ``postbasic calibrated data''-level images were downloaded from the Spitzer Heritage Archive\footnote{\url{https://sha.ipac.caltech.edu/applications/Spitzer/SHA/}} and processed through an automated image subtraction pipeline (for survey and pipeline details, see \citealp{kasliwal2017,jencson2019}). For reference images, we used the Super Mosaics,\footnote{Super Mosaics are available as Spitzer Enhanced Imaging Products through the NASA/IPAC Infrared Science Archive: \url{https://irsa.ipac.caltech.edu/data/SPITZER/Enhanced/SEIP/overview.html}} consisting of stacks of images obtained on 2005 February 2. Aperture photometry was performed on the difference images adopting the appropriate aperture corrections and Vega-system zeropoint fluxes from the IRAC instrument handbook\footnote{\url{http://irsa.ipac.caltech.edu/data/SPITZER/docs/irac/iracinstrumenthandbook/}} and following the method for a robust estimate of the photometric uncertainties as described in \citet{jencson2020_PhDT}. These data are presented in Figure \ref{fig:lc_all}.

\subsubsection{Spectroscopy}
We present 12 optical spectra of SN~2018lab ranging from less than 48 hours to over 300 days after explosion. Of the 12 spectra presented in this work, 11 were obtained as a result of a high-cadence spectroscopic follow up campaign using the Robert Stobie Spectrograph (RSS) on the Southern African Large Telescope \citep[SALT,][]{SALT} using a  1.50'' slit width, the FLOYDS instruments  \citep{LasCumbres} on the Las Cumbres Observatory's 2m Faulkes Telescopes North and South (FTN/FTS) with the set up described in \citet{LasCumbres} with a 2'' slit width, the Low Resolution Imaging Spectrometer \citep[LRIS,][]{LRIS} on Keck I using a 1.5'' slit width, and one of the Multi-Object Double Spectrographs \citep[MODS1,][]{MODS} on LBT in the 1.0'' segmented longslit configuration. The LBT spectrum from 308 days post-explosion is discussed in Section \ref{sec:nebspec}. We also include in our analysis the classification spectrum from 1.9 days post-explosion \citep{18labclass} taken as part of the Public European Southern Observatory (ESO) Spectroscopic Survey for Transient Objects \citep[ePESSTO,][]{ePESSTO} using the ESO Faint Object Spectrograph and Camera (EF0SC2) on the ESO New Technology Telescope (ESO-NTT) using a 1'' slit width with the Grism\#13 described in \citet{ePESSTO}. All spectra are logged in Table \ref{tab:specInst}.

\begin{table*}
 \centering
 \begin{tabular}{ c c c c c c}
    \hline
    Date & JD & Epoch (day) & Telescope & Instrument & Exposure (s)\\
    \hline
    2018-12-30 & 2458482.5411 & 1.6 & SALT & RSS & 1994.0\\ 
    2018-12-30 & 2458482.7795 & 1.9 & ESO-NTT & EFOSC2 & 600.0 \\ 
    2019-01-01 & 2458485.0255 & 4.1 & FTS 2m & FLOYDS & 3600.0 \\ 
    2019-01-04 & 2458487.9443 & 7.0 & FTN 2m & FLOYDS & 3600.0 \\
    2019-01-08 & 2458491.9115 & 11.0 & FTN 2m & FLOYDS & 3600.0 \\ 
    2019-01-11 & 2458494.8412 & 13.9 & Keck I & LRIS+LRISBLUE & 600.0\\ 
    2019-01-17 & 2458500.9915 & 20.1 & FTS 2m & FLOYDS & 3600.0 \\ 
    2019-03-02 & 2458544.8102 & 63.9 & FTN 2m & FLOYDS & 3600.0 \\ 
    2019-03-13 & 2458555.7320 & 74.8 & FTN 2m & FLOYDS & 3600.0 \\ 
    2019-03-28 & 2458570.7291 & 89.8 & FTN 2m & FLOYDS & 3600.0 \\ 
    2019-04-09 & 2458582.9357 & 102.0 & FTS 2m & FLOYDS & 3600.0 \\ 
    2019-11-01 & 2458788.9488 & 308.0 & LBT-SX & MODS1R & 600.0 \\
    \hline
 \end{tabular}
 \caption{Log of Spectroscopic Observations}
 \label{tab:specInst}
\end{table*}

\subsection{Distance}
We assume a distance modulus of 32.75$\pm{0.4}$ mag, based on the distance of 35.5 Mpc to IC2163/NGC2207 \citep{Theureau2007}.
This distance is a mean of the JHK Tully-Fisher distances and was used in \citet{Jencson2017}. This is consistent with the widely used distance to IC2163/NGC2207 of 35$\pm{2.5}$  Mpc \citep{Elmegreen2017,Kaufman2016} and the measured distance to NGC 2207, using SN Ia 1975A, of 39.6$\pm{5.5}$  Mpc \citep{Arnett1982}. A recent paper on SN~2010jp \citep{Corgan2022}, which is in the vicinity of IC2163/NGC2207, uses a distance of 24.5 Mpc. However, \citet{Corgan2022} also suggests that the host galaxy of SN~2010jp is a foreground dwarf galaxy, not IC 2163 or NGC 2207, which accounts for the difference in distances. 

\begin{figure}
    \centering
        \includegraphics[width=\hsize]{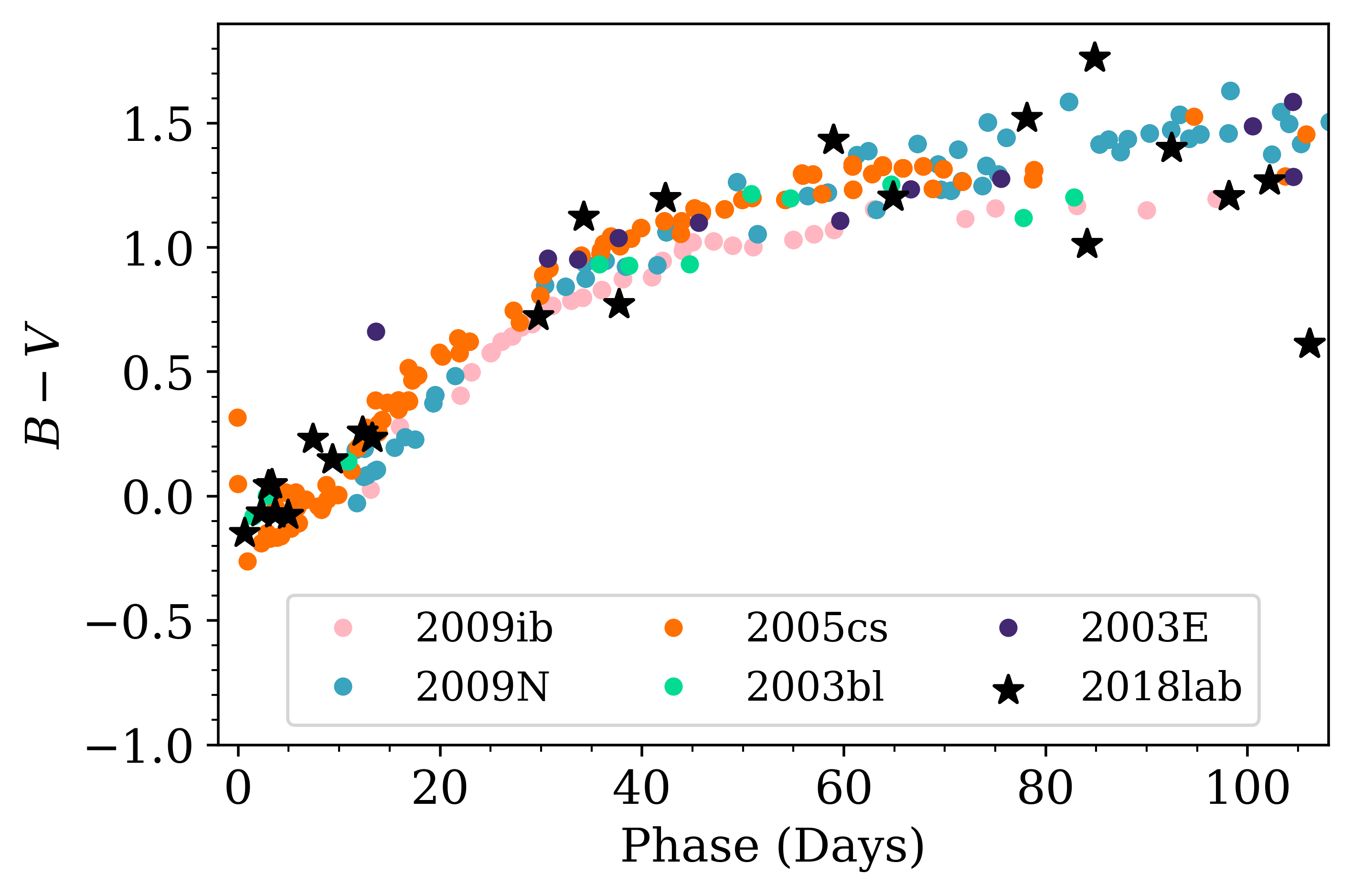}
    \caption{Extinction corrected $B-V$ color for SN~2018lab compared with other SNe with similar light curve properties. The adopted $E(B-V)_{tot} = 0.22$ is consistent with the color evolution of these similar SNe. Data shown for SN~2009ib \citep{09ib}, SN~2009N \citep{09N}, SN~2005cs \citep{05cs1, 05cs2, 05cs, 05cs4}, SN~2003bl \citep{Galbany2016}, SN~2003E \citep{Galbany2016}, SN~2003bl \citep{Anderson2014}, and SN~2003E \citep{Anderson2014}.}
    \label{fig:BV}
\end{figure}

\subsection{Extinction}
The equivalent widths of Na ID absorption lines correlate with interstellar dust extinction \citep{Richmond1994, Munari1997}. To estimate the extinction along the line of sight, the Na~ID features in the Keck LRIS spectrum, which has a high signal-to-noise and resolving power $R=715$, were analyzed. The equivalent widths of both the $z=0$ (Milky Way) and the $z=0.0089$ (host) features were measured by fitting and integrating Gaussian line profiles. The equivalent widths were then converted to $E(B-V)$ using Eq 9. of \citet{Poznanski2012} with an additional normalization factor of 0.86 from \citet{Schlafly2010}. This method gives a Milky Way extinction, $E(B-V)_\mathrm{MW} = 0.058_{-0.0095}^{+0.012}$ mag, which is roughly consistent with the value from \citet{Schlafly2011} of $0.0748\pm{0.0006}$ mag. We adopt the latter value. The equivalent width of the host Na ID doublet was close to 2~{\AA}. The relation between the Na ID equivalent width and dust extinction given in \citet{Poznanski2012} saturates at an equivalent width of $\sim$0.2{\AA}, so alternative methods of measuring SN~2018lab's host extinction are required.

The diffuse interstellar band (DIB) absorption feature used in \citet{Phillips2013} can also be used to determine extinction, however the DIB was not visible in any of the SN~2018lab spectra. This was also the case for the KI $\lambda$7699 line, which is effective at determining the host extinction as well \citep{Munari1997}. 

Host extinction is instead determined by comparing the color evolution of SN~2018lab to other SNe IIP with similar peak magnitudes and light curve shapes (light curve properties are described in Section \ref{sec:lcprop}), namely SN~2009ib \citep{09ib}, SN~2009N \citep{09N}, SN~2003bl \citep{Galbany2016, Anderson2014}, and SN~2003E \citep{Galbany2016, Anderson2014}. This analysis gives a $E(B-V)_\mathrm{host}$ of about 0.15 (see Figure \ref{fig:BV}). 
Using this value, the dereddened spectra of 2018lab matches the continuum slope of other extinction corrected LLSNe. Given the location of SN~2018lab in a dusty spiral arm of a star forming galaxy, this level of local host galaxy reddening is not surprising. 
There is likely significant uncertainty in this value, however the scatter in $B$$-$$V$ color makes it difficult to make a better estimation of the host extinction (Figure \ref{fig:BV}). We note that SN~2018lab exhibits evidence of CSM interaction, which can make a SN appear slightly more blue and may cause the extinction to be underestimated using this method. The combined Milky Way and host extinction gives a $E(B-V)_\mathrm{tot} = 0.22$ mag, which we adopt as the total extinction to the supernova.

\section{Photometric Evolution}\label{sec:lcprop}

\begin{figure}
    \centering
    \includegraphics[width=\hsize]{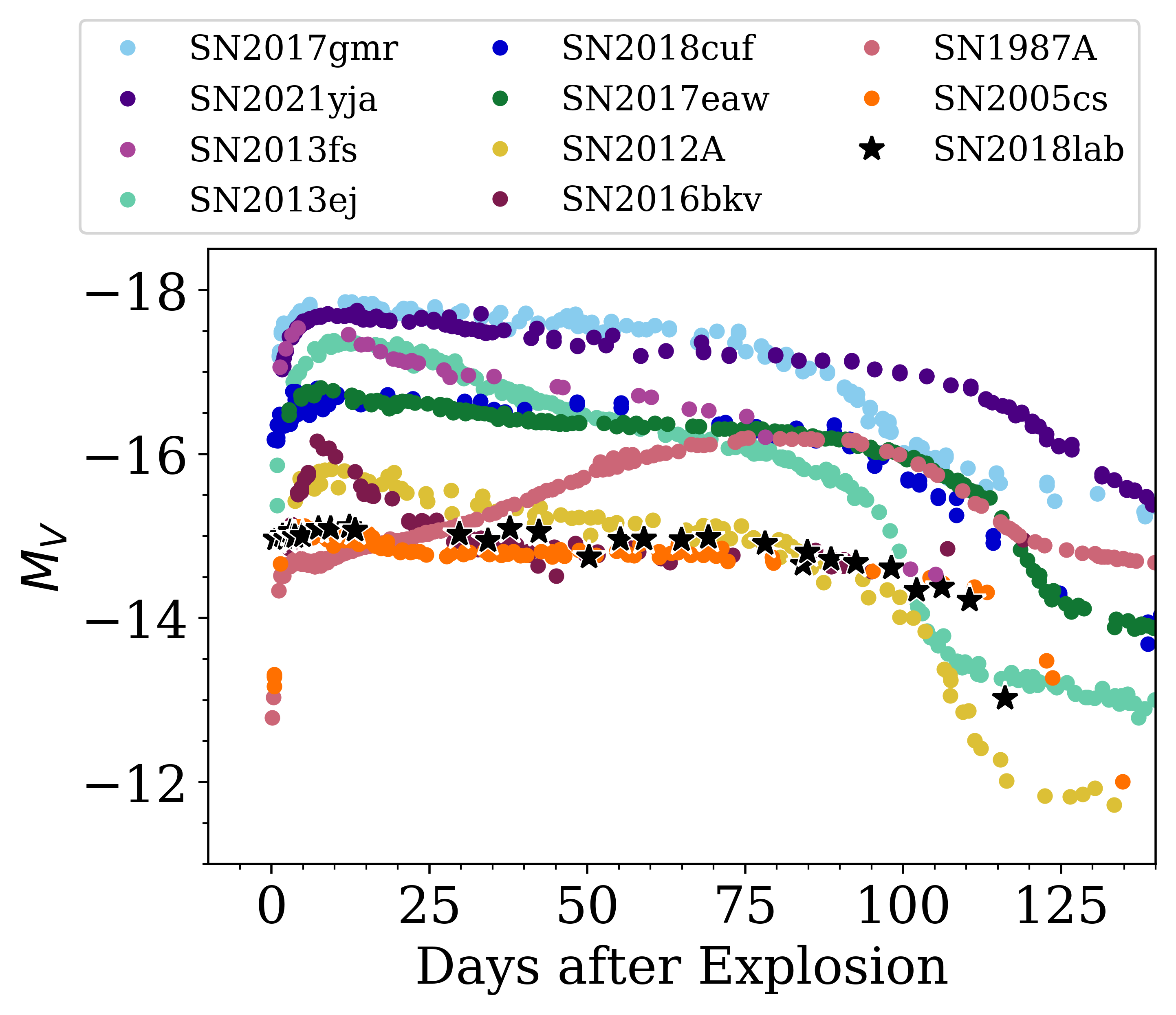}
    \caption{Absolute V-band light curve of SN~2018lab compared with other SNe II: SN~2017gmr \citep{Andrews2019}, SN~2021yja \citep{21yja}, SN~2013fs \citep{13fs1, Valenti2016, Bullivant2018}, SN~2013ej \citep{13ej1, 13ej2, 12A13ej}, SN~2018cuf \citep{18cuf}, SN~2017eaw \citep{17eaw, 17eaw2}, SN~2012A \citep{12A, 12A13ej}, SN~2016bkv \citep{16bkv, 16bkv2}, SN~1987A \citep{87A1,87A2,87A3,87A4}, and SN~2005cs \citep{05cs1, 05cs2, 05cs, 05cs4}. SN~2018lab has a peak V-band magnitude of $-$15.1 mag which is consistent with the luminosity of SN~2005cs, a notable LLSN.}
    \label{fig:V_lc}
\end{figure}

\begin{figure}
    \centering
    \includegraphics[width=\hsize]{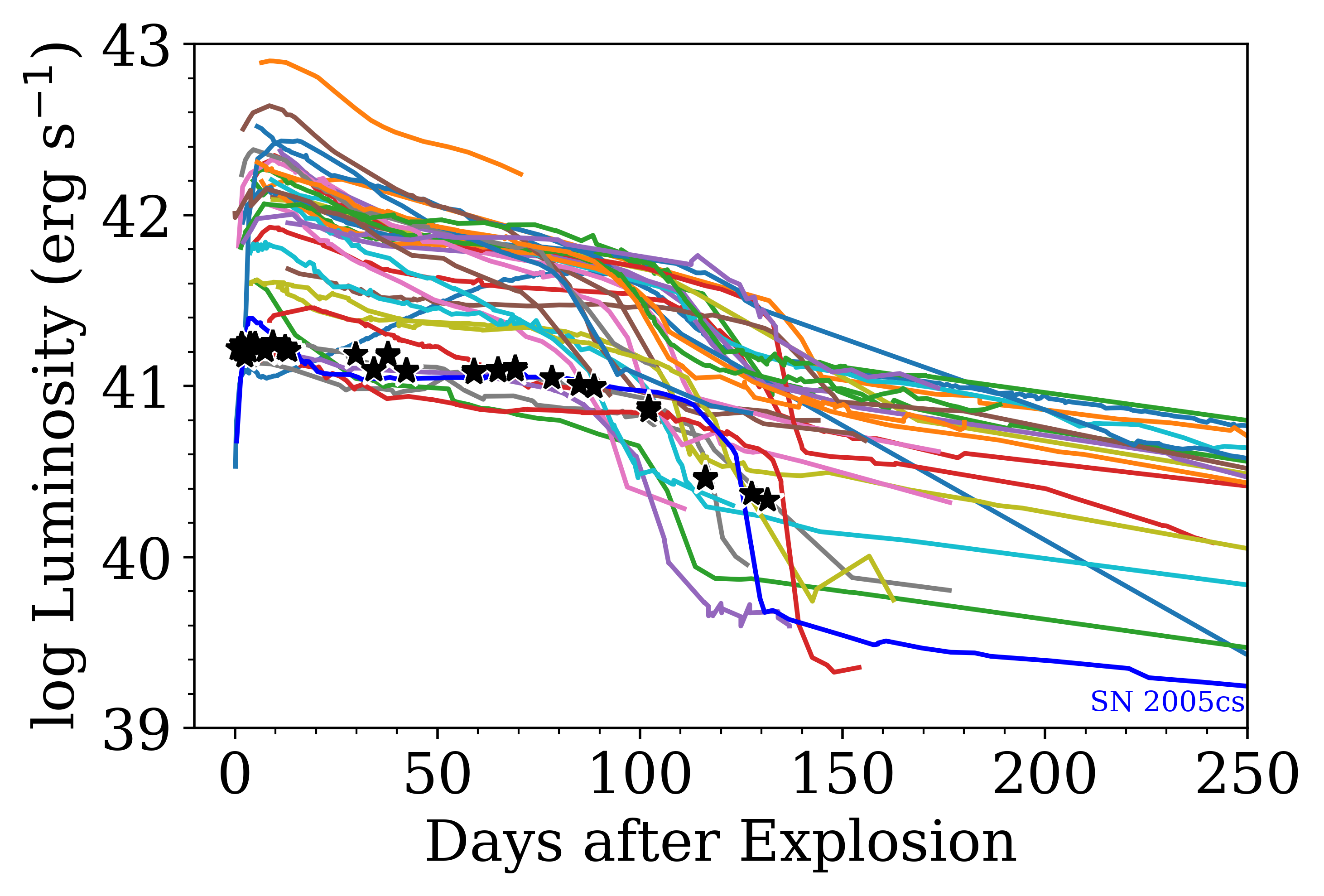}
    \caption{Bolometric light curve of SN~2018lab compared with other SNe II presented in \citet{Valenti2016}. SN~2018lab has a peak luminosity of 10$^{41.2 \pm 0.1}$ erg s$^{-1}$. While SN~2018lab has a flatter and shorter plateau than SN~2005cs (blue), these features are not atypical for a LLSN.}
    \label{fig:bolo_lc}
\end{figure}

In the V-band, SN~2018lab peaks at $-15.1\pm0.1$ mag, consistent with the observed brightness of the archetype LLSN SN~2005cs (see Figure \ref{fig:V_lc}).
Compared to SN~2005cs, the bolometric light curve of SN~2018lab remains fairly flat at the start of the plateau phase and has a shorter plateau duration. However, given a SN~2018lab peak luminosity of 10$^{41.2 \pm 0.1}$ erg s$^{-1}$, it fits well into the LLSN subclass, as shown in Figure \ref{fig:bolo_lc}.

The V-band decline rate of SN~2018lab in the 50 days following maximum brightness, denoted $s_{50}$, was measured according to the protocol outlined in \citet{Valenti2016}. SN~2018lab has an extremely flat plateau phase, with a $s_{50} = 0.13 \pm 0.05$ mag/50 days. There are very few light curve points at the end of the plateau making it difficult to fit the transition to the nickel tail, and therefore we are unable to estimate a reliable $^{56}$Ni mass.  The last few points of the r band light curve have a slope of $<0.01$ mag/day, indicating that they may lie on the nickel tail.
In order to get a rough estimate of the plateau length, we use the average time between the last point on the V band plateau and the first point on the tail in the r band to determine the plateau length $t_{PT} = 113 \pm 3$ days.

\citet{TESSdata} models the light curves of 20 CCSNe observed by \textit{TESS} including SN~2018lab (denoted as DLT18ar in their work) using a curved power-law (see their Eqn. 2). This method effectively reproduces the shape of SN~2018lab's early light curve, and they find a rise time of $t_\mathrm{rise} = 8.3 \pm 0.21$ days, which is among the fastest in their sample of 20 SNe. Additionally, SN~2018lab was the lowest luminosity SN in the sample by almost 2 magnitudes, with a peak luminosity of $-15.48\pm0.29$ mag in the TESS band. 

\begin{figure}
    \centering
    \includegraphics[width=\hsize]{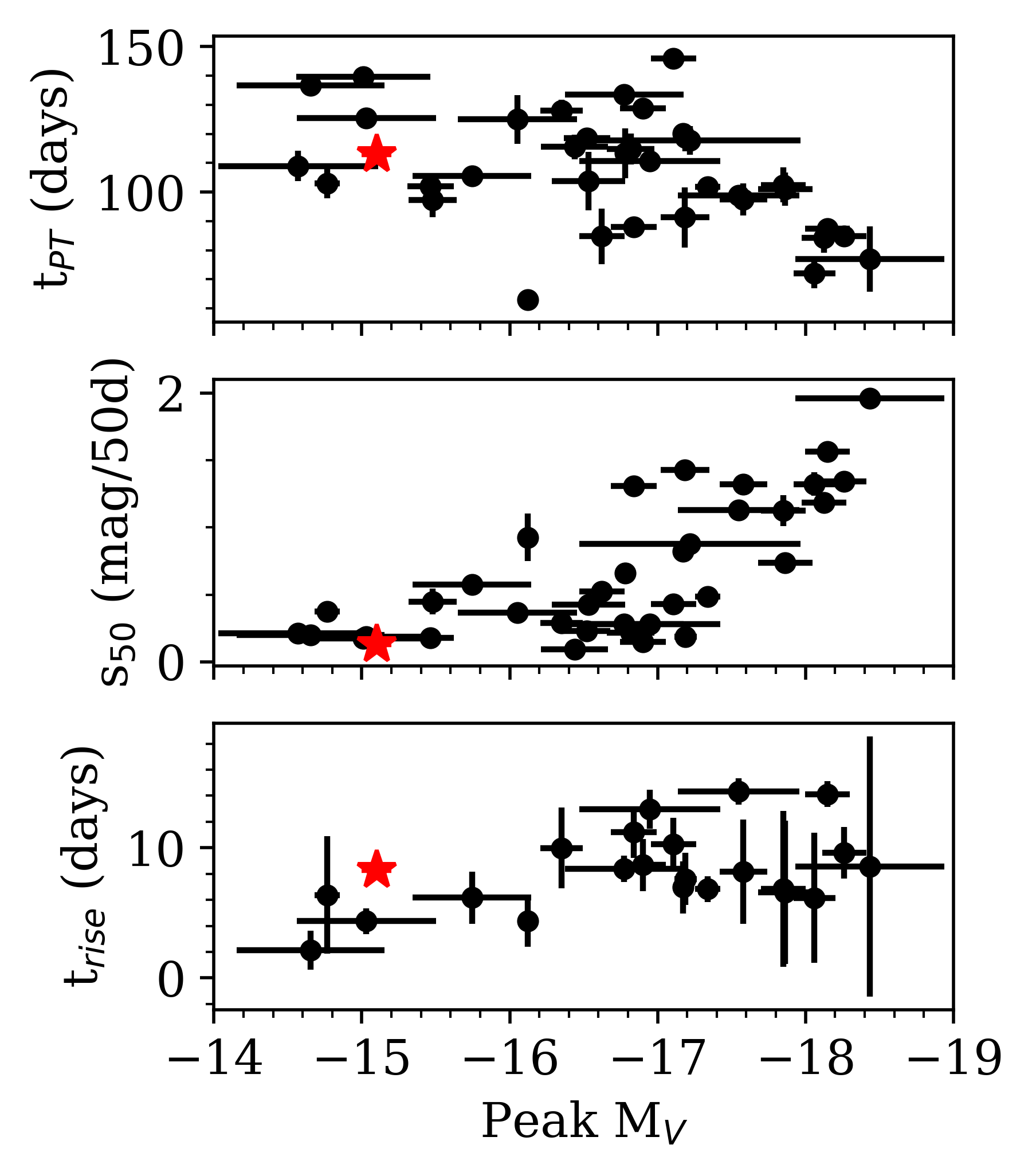}
    \caption{Comparison of the light curve properties of SN~2018lab to SNe in \citet{Valenti2016}. Top: SN~2018lab has a plateau length, $t_{PT}$, in agreement with other SNe II. Middle: In the 50 days following peak magnitude, the slope of SN~2018lab's light curve remains very flat, as is common in LLSNe. Bottom: Similar to other LLSNe, SN~2018lab rises to peak light quickly. Error bars on the SN~2018lab measurements are smaller than the marker.}
    \label{fig:lc_comp_param}
\end{figure}

The light curve properties of SN~2018lab (see Table \ref{tab:Parameters}) are in line with other LLSNe (see Figure \ref{fig:lc_comp_param}). 
The peak V-band luminosities of LLSNe, including SN~2018lab, are less than a typical CCSNe by a factor of 10 \citep{Pastorello2004}. 
The typical plateau time of SNe II, including LLSNe, is 80-140 days \citep{Valenti2016}, in agreement with SN~2018lab's $t_{PT}$. 
The peak luminosity and the decline rate of SNe II are related to one other, with LLSNe having much flatter plateaus (i.e. lower $s_{50}$ values) than more luminous SNe II \citep{Anderson2014}. The $s_{50}$ values for SNe II are $\lesssim$3 mag/50 days. Like other LLSNe, the $s_{50}$ of SN~2018lab lies on the low end of the $s_{50}$ continuum for SNe II. 
The rise times of SNe II are fast ($<$20 days) compared to other types of SNe; the rise times of LLSNe are on the faster end of the SNe II distribution with $t_{rise} \lesssim 10$ days \citep{Valenti2016}.
The values of $s_{50}$, $t_{PT}$, and $t_{rise}$ for SN~2018lab are similar to  other LLSNe in \citet{Valenti2016}.

\begin{table}
 \centering
 \caption{SN~2018lab Parameters} \label{tab:Parameters}
 \begin{tabular}{ c c c c c c}
    \hline
    Last Non-Detection & JD 2458480.6624\\ 
    Discovery & JD 2458481.626 \\
    Explosion Epoch\footnote{taken from \citet{TESSdata}\label{Vallely}} &  JD 2458480.9 $\pm$ 0.1 \\
    Redshift $z$ & 0.0089 $\pm$ 0.0001 \\
    Distance (modulus $\mu$) & 35.5 Mpc (32.75 mag) \\
    $E(B-V)_\mathrm{tot}$ & 0.22 mag \\
    $M_\mathrm{TESS}$ at peak\footref{Vallely} & $-15.48$ $\pm$ 0.29 mag \\
    $t_{rise}$\footref{Vallely} & 8.3 $\pm$ 0.21 days \\
    $s_{50}$\footnote{as defined by \citet{Valenti2016}} & 0.13 $\pm$ 0.05 mag/50 days \\
    $t_{PT}$ & 113 $\pm$ 3 days \\
    \hline
 \end{tabular}

\end{table}

\subsection{Shock Cooling Model}
The rising light curves of SNe II are in part powered by shock cooling---energy added to the stellar envelope by the core-collapse shock wave. To determine the effect of shock cooling on the rising light curve of SN~2018lab, the light curve is fit using the Light Curve Fitting package \citep{lightcurvefitting} which employs the analytic method for modeling early SNe II light curves powered by shock cooling described in \citet{Sapir2017}. 

\begin{figure*}
    \centering
    \includegraphics[width=\hsize*9/10]{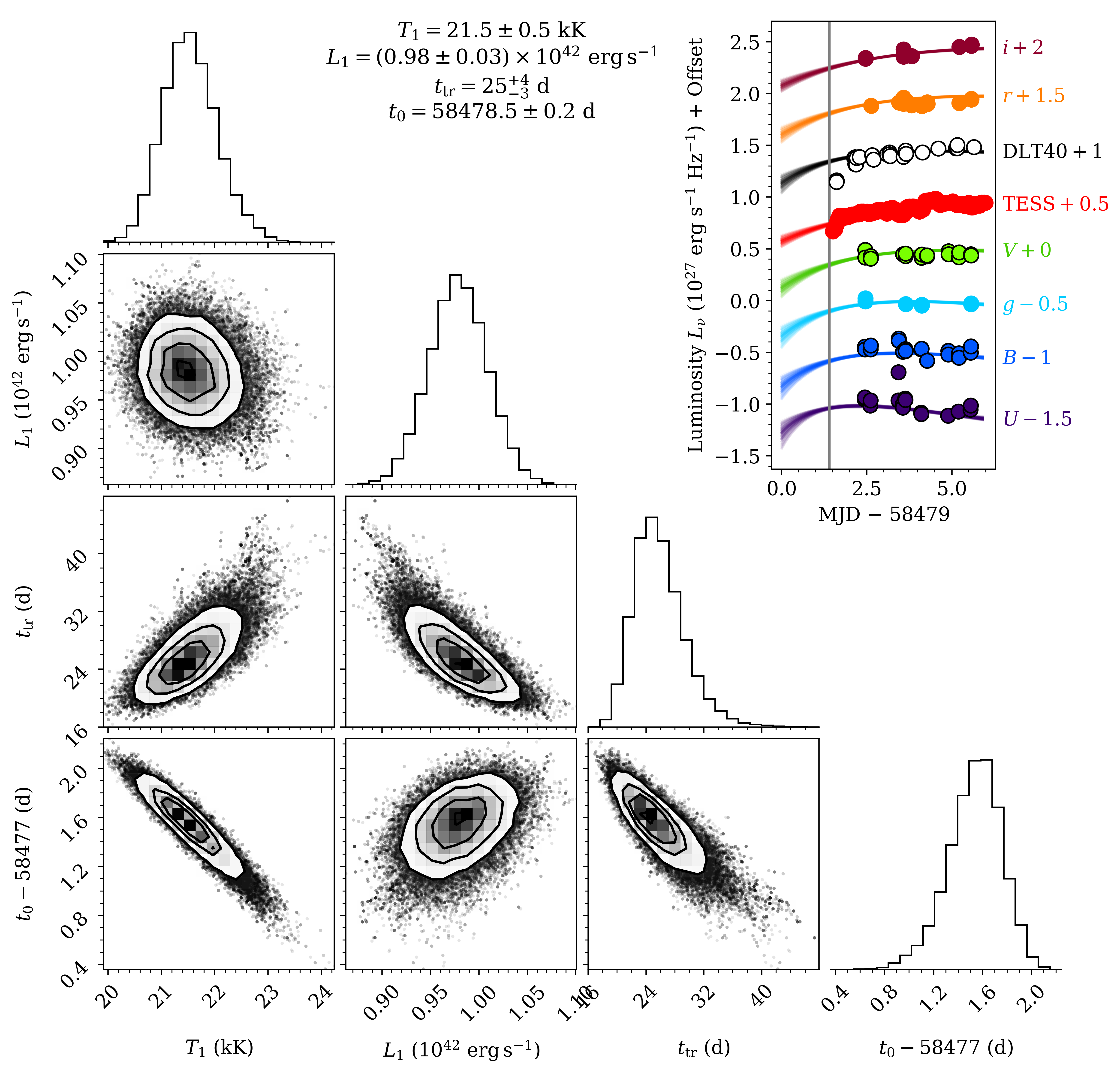}
    \caption{Posterior distributions of and correlations between the temperature 1 day after explosion ($T_1$), the total luminosity $\sim$1 day after explosion ($L_1$), the time at which the envelope becomes transparent ($t_{tr}$), and the explosion time ($t_0$). The 1$\sigma$ credible intervals, centered about the median, are given at the top. The top right panel shows 100 randomly drawn fits from the MCMC compared to the multiband light curve. The fit predicts an explosion time significantly ($>1$ day) before the explosion epoch (vertical grey line) constrained by TESS observations \citep{TESSdata} and DLT40 non-detections. The fit fails to describe SN~2018lab's fast rising light curve (see the TESS and DLT40 light curves). This may indicate the presence of ejecta-CSM interaction, which is not accounted for in this model.}
    \label{fig:ShockCooling}
\end{figure*}
 
Degeneracies between the Sapir \& Waxman model parameters makes it difficult to fit them independently in the case of SN~2018lab. Therefore, we use the version of the Sapir \& Waxman model used in \citet{16bkv} which utilizes scaling parameters: the temperature 1 day after explosion ($T_1$), the total luminosity $\sim$1 day after explosion ($L_1$), the time at which the envelope becomes transparent ($t_{tr}$), and the explosion time ($t_0$).
This version of the model, with a polytropic index n=1.5 for a RSG progenitor density profile, was fit to the multiband light curve of SN~2018lab up to MJD 58485 (4.6 days after explosion). 
This was done with a Markov Chain Monte Carlo (MCMC) routine and flat priors for all parameters. The model gives the total luminosity and blackbody temperature as a function of time for each set of parameters. This is then converted to observed fluxes for each photometry point. Figure \ref{fig:ShockCooling} shows the results of the MCMC, including the light curve fits, posterior distributions, and the 1$\sigma$ credible intervals centered on the medians.

The best fit models have difficulty reproducing the fast rise, completely missing the DLT40 and TESS rise points. The best fit explosion time is MJD 58478.5$\pm0.2$, $>$1 day before the highly constrained explosion time estimated from the TESS data \citep[MJD 58480.4$\pm$0.1,][]{TESSdata} and before two DLT40 non-detections. Further, the model fails to fit the rising light curve when the explosion time is fixed to be within the error of the TESS explosion epoch. Due to the failure of the model to accurately fit the steep rise in the light curve, we do not consider these models to be a good fit, but they are included here for completeness. 

The failure of the shock cooling model to accurately predict the steep rise may be evidence of ejecta-CSM interaction, which is not accounted for in the Sapir \& Waxman model. A steep rise can occur when the CSM is optically thick enough that shock breakout does not occur on the edge of the stellar envelope but rather outside of it, within the CSM. The gradual density gradient of the CSM means this shock breakout occurs at a lower density than for a bare RSG, allowing the shocked material to cool and expand faster, resulting in early excess flux, and therefore a steeper rise than would be expected for a SN without CSM \citep{Morozova2017, Tinyanont2022}. This explanation is bolstered by the presence of broad-lined flash features in the early spectra ($<2$ days post-explosion), to be discussed in Section \ref{sec:EarlySpec}. 

\begin{figure*}
    \centering
    \includegraphics[width=\hsize]{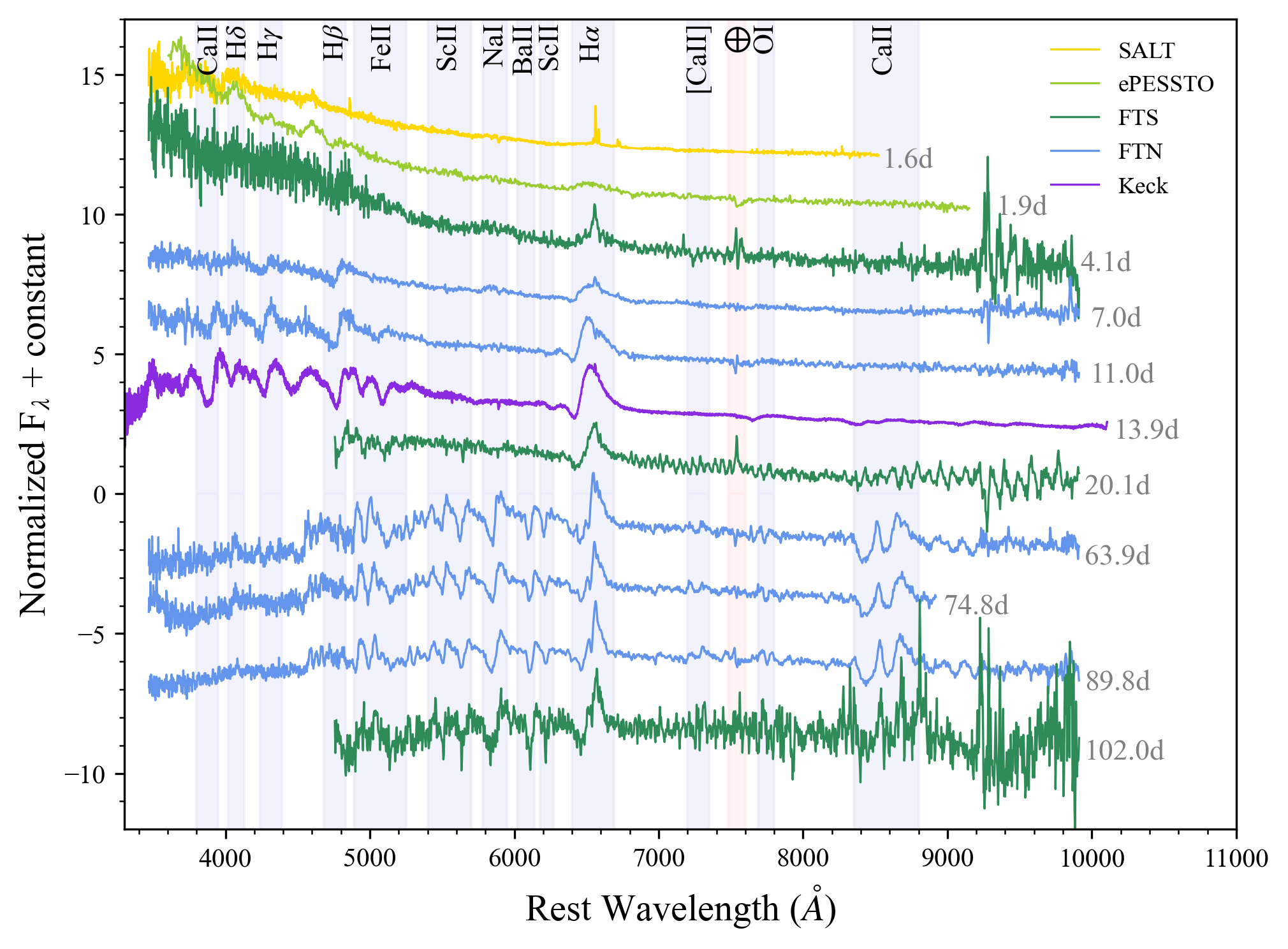}
    \caption{Spectroscopic evolution of SN~2018lab, corrected for $E(B-V)_\mathrm{tot} = 0.22$. The early spectra ($\le$7 days) of SN~2018lab are dominated by blue continua and H and He~I emission lines. At 11 days, Ca~II H\&K and Fe~II lines begin to appear. At later times ($>$50 days), more metal lines begin to appear: O~I, Ca~II infrared triplet, [Ca~II], and Na~ID. Strong Sc~II and Ba~II lines also emerge during this epoch. All of these spectral features and the timing of their emergence are typical of LLSNe \citep{Pastorello2004, Spiro2014, Gutierrez2017}.}
    \label{fig:spec}
\end{figure*}

\section{Spectral Evolution}\label{sec:Spec}

The spectra $<$105 days post explosion of SN~2018lab are presented in Figure \ref{fig:spec}. Based on the 2D spectra, we attribute the narrow lines, particularly near H$\alpha$, to host contamination. While there could be narrow lines from the SN, we are unable to identify them given the nearby H~II region.

The spectral evolution of SN~2018lab is similar to that of other LLSNe presented in previous papers \citep[e.g.][]{Benetti2001, Pastorello2004, 05cs, Spiro2014, 09N, 08bk, 21aai}.
The first 4 spectra ($\le$7 days) exhibit a blue continuum and the slow emergence of Balmer lines and He~I $\lambda$5876, as is typical of all SNe II. These early lines have P Cygni profiles with very shallow absorption components. In the 11 day spectrum, the Ca~II H\&K ($\lambda$3934, $\lambda$3968) and the Fe~II multiplet 42 ($\lambda$4924, $\lambda$5018, $\lambda$5169) lines become visible while He~I $\lambda$5876 disappears. In the second half of the plateau ($>$50 days), the O~I $\lambda$7774, Ca~II infrared triplet ($\lambda$8498, $\lambda$8542, $\lambda$8662), [Ca~II] ($\lambda$7291, $\lambda$7324), and Na~ID ($\lambda$5890, $\lambda$5896) lines appear and strengthen. Further, this epoch also exhibits the characteristic strong Sc~II and Ba~II lines seen in LLSNe \citep{Pastorello2004, Spiro2014, Gutierrez2017}.

There are a few notable features in the spectral evolution of SN~2018lab worth further discussion: the broad-lined flash feature in the early spectra, the appearance of an additional
absorption component on the blue side of H$\alpha$, and the evolution of the H$\alpha$ profile in the second half of the plateau phase. These features are discussed in sections \ref{sec:EarlySpec}, \ref{sec:Cachito}, and \ref{sec:multiH} respectively.

\subsection{Flash Spectroscopy}\label{sec:EarlySpec}

\begin{figure}
    \centering
    \includegraphics[width=\hsize]{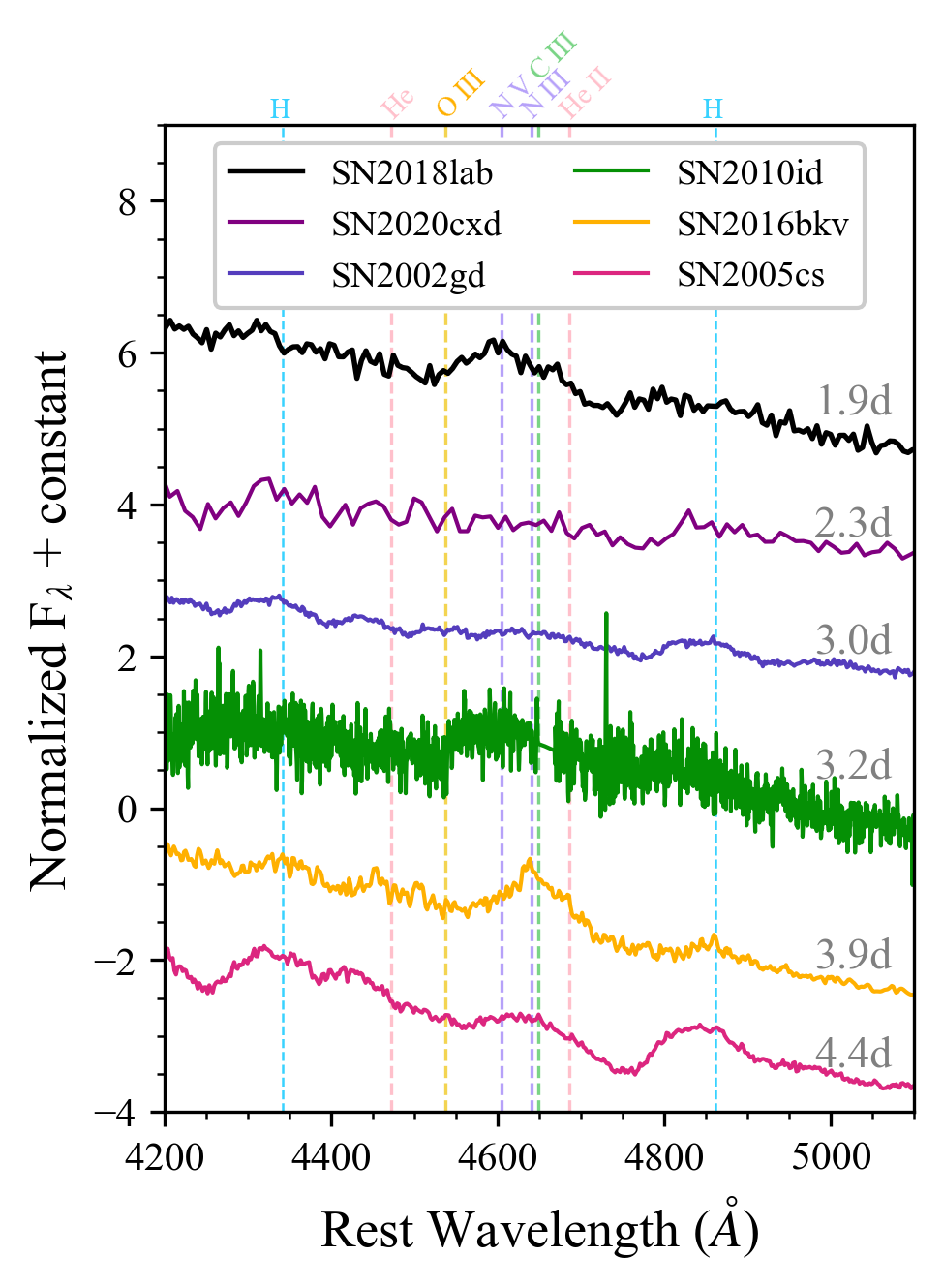}
    \caption{Comparison of the broad early feature seen in SN~2018lab with other LLSNe which have spectra $<$5 days post-explosion: SN~2002gd \citep{Spiro2014}, SN~2005cs \citep{05cs5,05cs}, SN~2010id \citep{10id}, SN~2016bkv \citep{16bkv}, and SN~2020cxd \citep{21aai}. Of these LLSNe, 4 (SN~2005cs, SN~2010id, SN~2016bkv, and SN~2018lab) out of the 6 have a strong broad early spectral feature near 4600 \AA, though the shape, strength, and interpretation of this feature varies. It is likely that more LLSNe also exhibit similar features however a larger sample of early time ($<$5 days) spectra is needed to constrain the frequency at which this feature occurs. All spectra are extinction corrected.}
    \label{fig:LLSNeComp}
\end{figure}

SN~2018lab does not exhibit narrow high-ionization lines in the early ($<$2 days) spectra. Instead, early spectra of SN~2018lab show a broad feature from 4500 to 4750 {\AA} (see Figure \ref{fig:LLSNeComp}). This feature peaks near the N V $\lambda$4604 line. The feature is most clear in the spectrum 1.9 days post-explosion though it is also present in the first spectrum of SN~2018lab (1.6 days post-explosion). The SN~2018lab spectrum from 4.1 days post-explosion has low signal-to-noise in the relevant wavelength range and we are unable to discern if the earlier broad feature remains. Only one LLSNe has exhibited narrow high-ionization lines, SN~2016bkv \citep{16bkv}. In the spectra of SN~2016bkv, broad-lined flash features first appear in the spectra taken 4 days post-explosion in a shape similar to those seen in SN~2018lab, and the narrow lines become prominent a day later.

\begin{figure}
    \centering
    \includegraphics[width=\hsize]{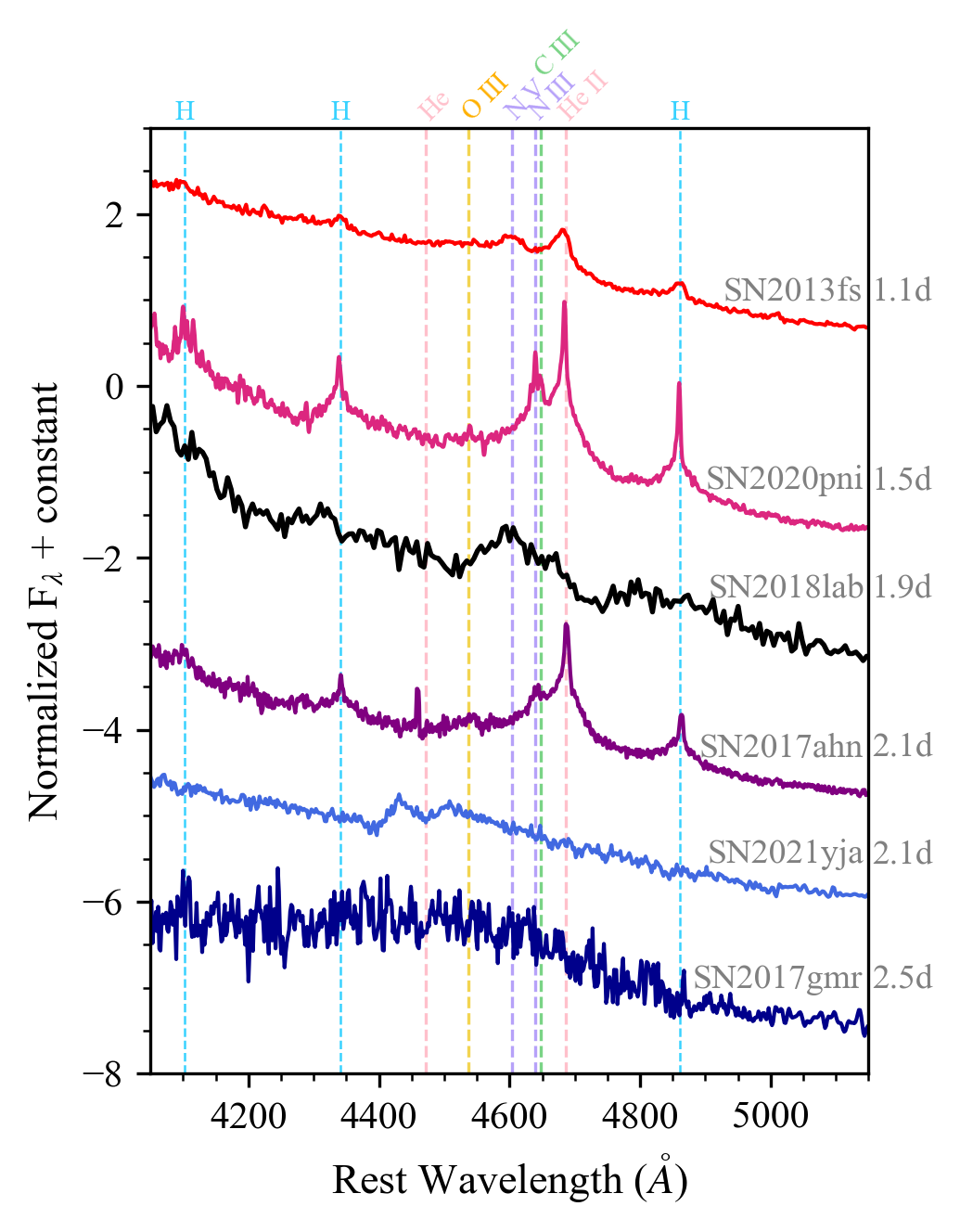}
    \caption{The 1.9 day spectrum of SN~2018lab compared to early time spectra of SN~2013fs \citep{Bullivant2018}, SN~2020pni \citep{20pni}, SN~2017ahn \citep{17ahn}, SN~2021yja \citep{21yja}, and SN~2017gmr \citep{Andrews2019}. SN~2020pni and SN~2017ahn exhibit clear narrow-lined flash features; where as SN~2013fs, SN~2021yja, and SN~2017gmr all have broader early spectral features. The morphology of the feature in the spectra of SN~2018lab further highlights the significant diversity of flash spectroscopy observed in SNe II. All spectra have been corrected for extinction.}
    \label{fig:FlashCompare}
\end{figure}

An early broad feature near 4600 {\AA}, sometimes referred to as a ``ledge" feature \citep{Andrews2019, Soumagnac2020, 21yja}, has been observed in the early spectra of other SNe II (see Figure \ref{fig:LLSNeComp} and \ref{fig:FlashCompare}). Very few LLSNe have spectra  $<$5 days following explosion. However of those that do---SN~2002gd \citep{Spiro2014}, SN~2005cs \citep{05cs5}, SN~2010id \citep{10id}, SN~2016bkv \citep{16bkv}, and SN~2020cxd \citep{21aai}---the majority (SN~2005cs, SN~2010id, SN~2016bkv) appear to have a feature similar to what we observe for SN~2018lab (see Figure \ref{fig:LLSNeComp}). The cause of this feature has been explained in three ways. In the spectra of SN~2005cs, \citet[][their Fig. 5]{05cs5} interprets this feature as high velocity (HV) H$\beta$. There is no indication of a HV feature blueward of H$\alpha$ in SN~2018lab at early times, so we disfavor this explanation. An alternative explanation is provided for SN~2010id by \citet[][their Fig. 2]{10id}, which suggests that this feature is broad, blue-shifted He~II $\lambda$4686. This analysis has been used to explain similar features in more typical SNe II as well, as seen in \citet[][their Fig. 10]{Quimby2007}, \citet[][their Fig. 20]{Bullivant2018}, and \citet[][their Fig. 18]{Andrews2019}. The other interpretation is that the feature is the blend of several ionized features from the CSM \citep{Dessart}. This is the explanation used by \citet[][their Fig. 2]{16bkv} to explain the shape of the feature in the spectra of SN~2016bkv, and has also explained similar features in more typical SNe II, as seen in \citet[][their Fig. 7]{Soumagnac2020}, \citet[][their Fig. 5]{Bruch2021}, and \citet[][their Fig. 11]{21yja}. SN~2018lab's early broad feature is somewhat double peaked indicating that there may be more than one line contributing to the feature. Therefore we posit that this feature is likely the blend of several ionized features from the CSM: N~V, N~III, C~III, O~III, and He~II, rather than just blue-shifted He~II (see Figure~\ref{fig:LLSNeComp}). 

The morphology of SN~2018lab's ledge feature adds to the significant diversity observed in the early spectra of SNe II (see Figure \ref{fig:FlashCompare}).
Symmetric narrow-lined flash features, like those seen in SN~2017ahn \citep{17ahn} and SN~2020pni \citep{20pni} are produced via non-coherent scattering of thermal electrons. In contrast, bulk motions produce broad lines which can blend together and produce a broad asymmetric feature \citep{Dessart2009}. When observed, both narrow- and broad-lined flash features can be used as a probe of the properties of the progenitor and the extent of the CSM. 

\begin{figure}
    \centering
    \includegraphics[width=\hsize]{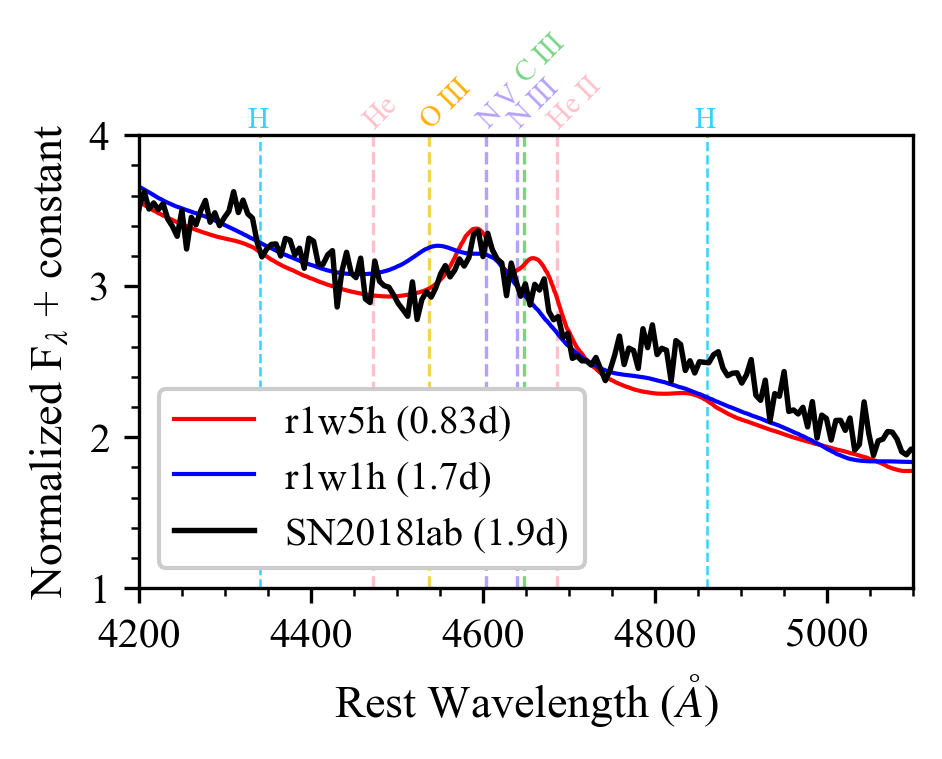}
    \caption{The early broad feature (4500-4750 \AA) in the 1.9 day spectrum of SN~2018lab is compared to the \citet{Dessart} models that most closely resemble the spectra, \texttt{r1w1h} and \texttt{r1w5h}, which are both RSGs with extended atmospheres. The closest analogue to the SN~2018lab data is the the \texttt{r1w5h} model at 0.83 days. \texttt{r1w1h} and \texttt{r1w5h} are scaled by 70\% to better fit the flux of the feature and have been convolved to the resolution of the observed spectrum \citep[14 \AA,][]{ePESSTO}.}
    \label{fig:Dessart}
\end{figure}

As shown in Figure \ref{fig:Dessart}, the broad early spectral feature in the 1.9 day spectrum of SN~2018lab closely resembles the \citet{Dessart} \texttt{r1w1h} and \texttt{r1w5h} models, both of which have RSG progenitors  with extended atmospheres and CSM.
The correspondence with the \texttt{r1w5h} model is especially striking.
The \citet{Dessart} r1w1 and r2w1 models also display ledge features, however these features are blue-shifted with respect to the observed SN~2018lab feature and are therefore not included in Figure \ref{fig:Dessart}.

Both the r1w1h and r1w5h models display narrow-lined flash features which appear immediately following explosion ($<$4 hrs) and quickly evolve into a broad spectral feature. 
These models focus on the first $\sim$15 days after explosion and only extend out to $1.5\times10^{15}$ cm.
Both \texttt{r1w1h} and \texttt{r1w5h} assume a progenitor star with a radius $R_* = 501 R_\odot$ and a wind mass loss rate of $10^{-6}$ and $5\times10^{-3}$ $M_\odot \ \mathrm{yr}^{-1}$, respectively. Both have extended atmospheres, with scale heights of $H_\rho = 0.3 R_*$ for \texttt{r1w1h} and $H_\rho = 0.1 R_*$ for \texttt{r1w5h}.
A moderate amount of energy deposited into an RSG envelope in late-stage nuclear burning can cause envelope expansion and mass ejection \citep{Smith2014, Morozova2020}. Just like dense CSM, an extended envelope can produce excess luminosity in SNe light curves \citep{Morozova2020}. 
The shape of SN~2018lab's early broad feature is qualitatively reproduced by the \texttt{r1w1h} and \texttt{r1w5h} models. Note that these models assume a much more energetic explosion ($1.35\times10^{51}$ ergs) and a much more massive progenitor (ejecta mass of 12.52 $M_\odot$) than is typical for LLSNe, therefore the CSM around SN~2018lab is unlikely to have identical properties to the modeled CSM. However, the similarity of the observed ledge feature to that of the r1w1h and r1w5h models could indicate that the feature may be caused by an extended envelope of an RSG progenitor and CSM interaction.

The ledge feature seen in the SN~2018lab data is most similar to \texttt{r1w5h} at 0.83 days. The similarity to the \texttt{r1w5h} model suggests the presence of a higher density CSM than assumed by the \texttt{r1w1h} model, but still low enough to prevent the appearance of narrow-lined flash features more than a few hours after explosion. The early broad-lined flash features in the spectra of SN~2016bkv are also similar to the shape of the \texttt{r1w5h} model at 0.83 days. However, this spectral feature in SN~2016bkv appears 4 days post-explosion, substantially after the model epoch, which may suggest a much larger and denser CSM than described by the model \citep{18zdEC}. In SN~2018lab, the features are present much earlier, indicating a progenitor with an extended envelope similar to that described by the \texttt{r1w5h} model with less dense CSM than SN~2016bkv.

\subsection{Cachito Features}\label{sec:Cachito}

\begin{figure}
    \centering
    \includegraphics[width=\hsize]{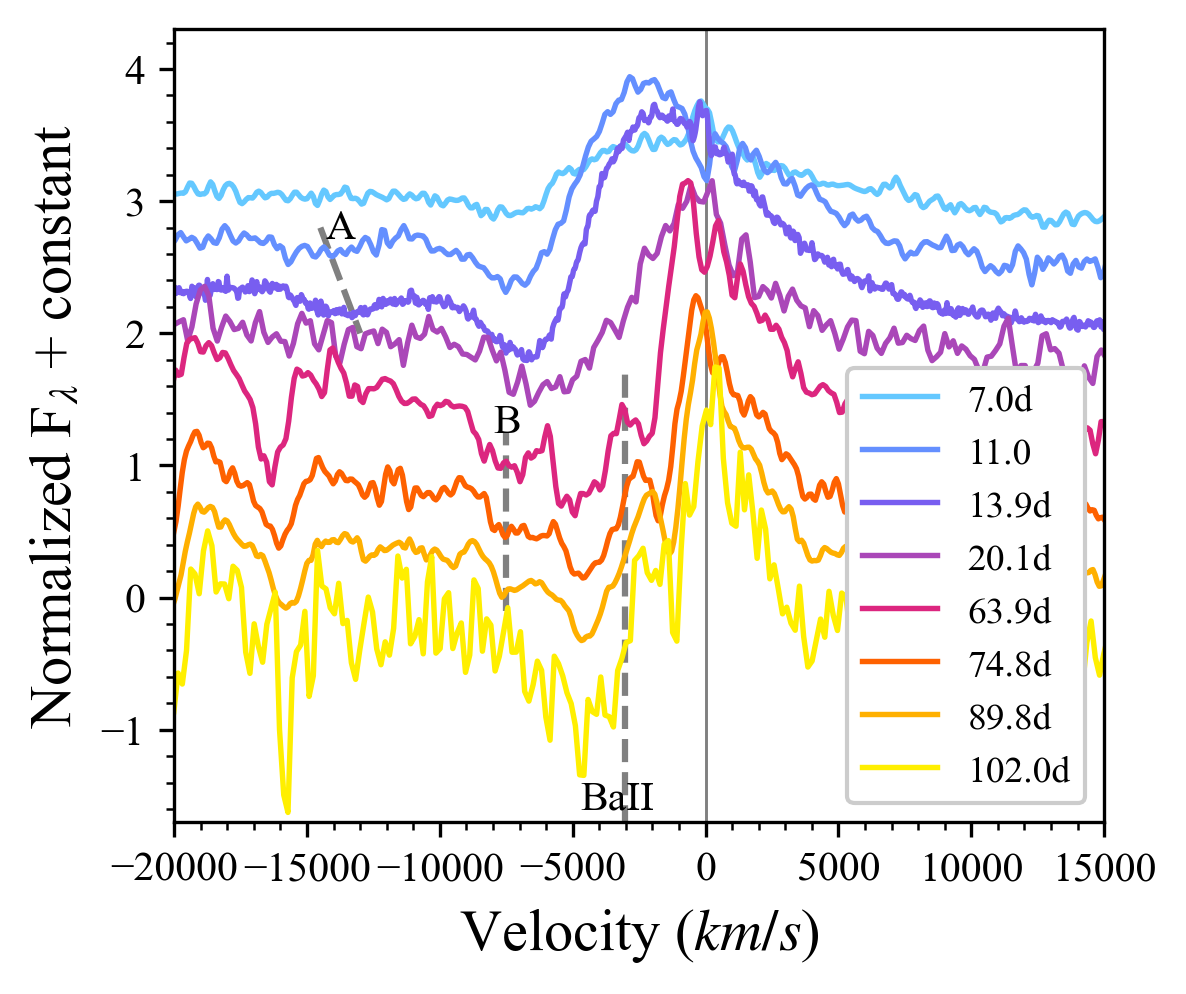}
    \caption{Evolution of H$\alpha$ starting at 7 days post-explosion. The spectra exhibit both A (10-15 days) and B (50-90 days) Cachito features (dotted lines). These features are likely due to Si~II $\lambda6355$ and HV~H respectively. At $>50$ days the existence of a complex H$\alpha$ profile becomes evident. This is attributed to the presence of Ba~II $\lambda6497$ (rest-frame denoted with green dashed line).}
    \label{fig:Halpha_evol}
\end{figure}

``Cachito" features \citep{Gutierrez2017} are small absorption features blueward of H$\alpha$ which are common in the optical spectra of SNe II \citep[e.g.][]{15oz, 18ivc, 18cuf}.  There are two main types of Cachito features, the kind which arise earlier ($<$40 days) in the spectral evolution and those which emerge later ($>$40 days). 
Both types of Cachito feature appear on the blue side of H$\alpha$ in the spectra of SN~2018lab, and are distinct (see Figure \ref{fig:Halpha_evol}). 
\citet{Gutierrez2017} found that, among SNe that exhibit Cachito features at $<$40 days post-explosion, in 60\% of cases the feature results from Si~II $\lambda6355$ and the remaining cases are likely due to high velocity (HV) H$\alpha$. In SNe with Cachito features that emerge at $>$40 days, this feature may occur when X-rays from the SN shock ionize and excite the outer unshocked ejecta and HV~H absorption forms \citep[see][]{Chugai2007}.
 
The early Cachito feature, denoted as A in Figure \ref{fig:Halpha_evol}, appears in the 11 and 13.9 day spectra at $13000-14000$ km s$^{-1}$ with respect to rest H$\alpha$. If the `A' Cachito feature is due to Si~II $\lambda6355$ it should have a velocity similar to other metal lines in the spectrum \citep{Gutierrez2017}. The measured velocity of the shallow `A' Cachito feature in the 13.9 day spectrum of SN~2018lab is 4500 km s$^{-1}$ in the Si~II  $\lambda6355$ rest frame. This velocity is similar to the velocity of Fe~II $\lambda5018$ and $\lambda5169$ in the same epoch.
We determine that the Cachito feature in the 11 and 13.9 day spectra of SN~2018lab is likely the result of Si~II $\lambda6355$. 

The late Cachito feature, denoted as B in Figure \ref{fig:Halpha_evol}, appears in the spectra from 50-90 days post explosion. While Ba~II $\lambda6497$ is visible in this region during the relevant epochs, a velocity analysis indicates that the `B' Cachito feature in SN~2018lab is not associated with Si~II $\lambda6355$ or Ba~II $\lambda6497$. If the `B' Cachito is related to HV~H, its velocity should be similar to that of H$\alpha$ at earlier phases and a companion feature may be visible blueward of H$\beta$, though this is rare in the LLSNe subclass \citep{Gutierrez2017}. 
The velocity relative to H$\alpha$ of the `B' Cachito feature, $\sim$7500 km s$^{-1}$, is consistent with the velocity of H$\alpha$ in the 7 and 11 day spectra.
This indicates that the Cachito feature in the 50-90 day spectra of SN~2018lab is likely the result of HV~H.
The numerous metal lines and low signal-to-noise on the blue end of the spectra make it difficult to discern if there is a counterpart HV feature near H$\beta$. This HV~H feature is likely to be related to SN ejecta and RSG wind interaction \citep{Gutierrez2017} and may be further evidence for CSM surrounding the progenitor.

\begin{figure}
    \centering
    \includegraphics[width=\hsize]{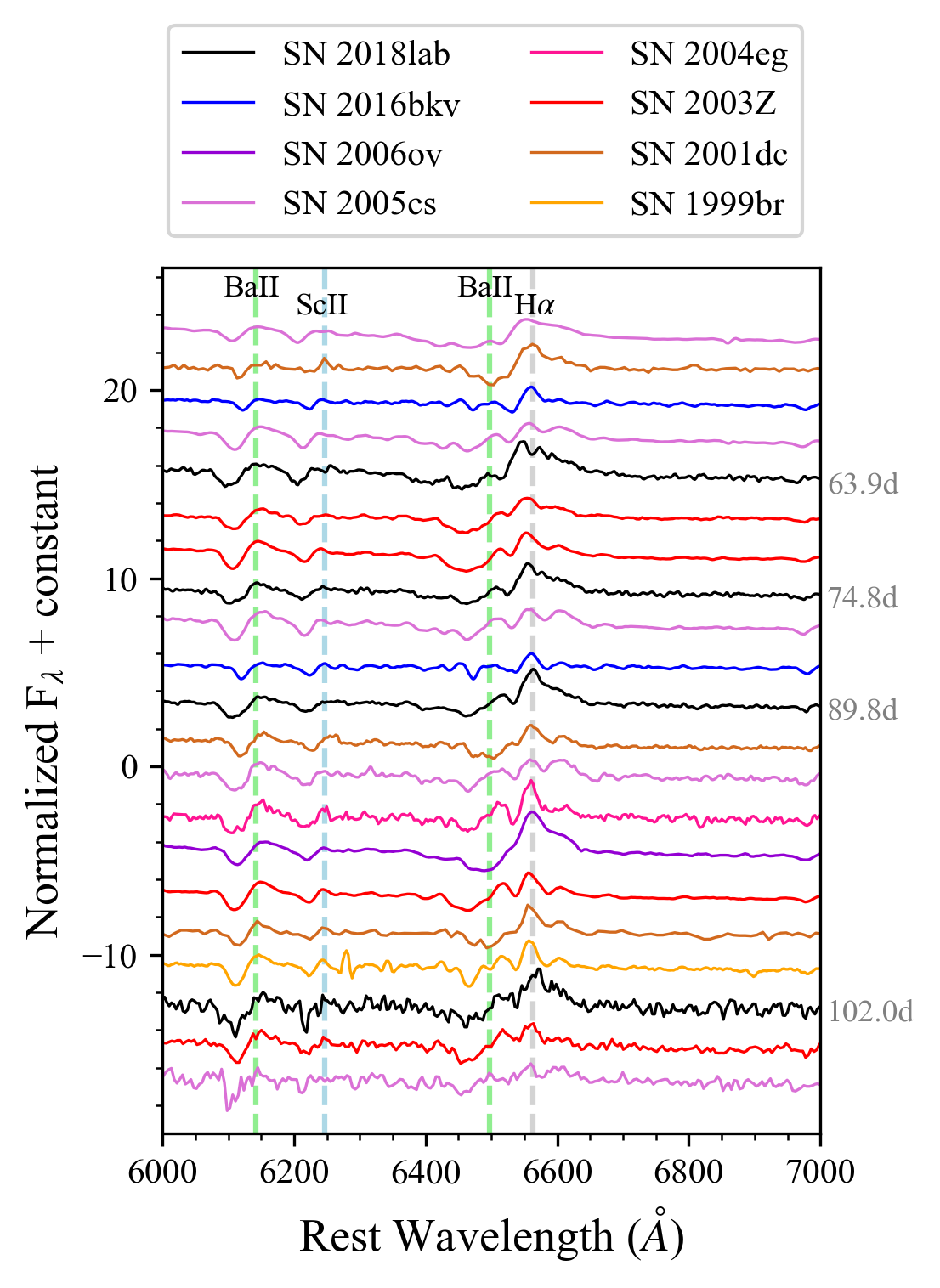}
    \caption{Spectra from the later half of the plateau phase of other notable LLSNe compared to SN~2018lab. Many LLSNe have strong Ba~II and Sc~II lines during this phase of evolution which yields a complex H$\alpha$ profile. Data taken from \citet[][SN~1999br and SN~2001dc]{Pastorello2004}, \citet[][SN~2005cs]{05cs}, \citet[][SN~2003Z, SN~2004eg, and SN~2006ov]{Spiro2014}, \citet[][SN~2016bkv]{16bkv}. All spectra are corrected for extinction.} 
    \label{fig:LLSNe_HaComp}
\end{figure}

\subsection{Complex H$\alpha$ Profile}\label{sec:multiH}
 
The H$\alpha$ in SN~2018lab exhibits a clear P Cyngi profile beginning at the start of the plateau phase. In the spectra taken $7-20$ days post-explosion, the H$\alpha$ velocity is $6000-8000$ km s$^{-1}$. This is similar to the H$\alpha$ velocities observed for SN~2005cs at the same epochs \citep{05cs}. CSM interaction will decelerate SN ejecta, with high density CSM resulting in ejecta speeds $\sim$1000 km s$^{-1}$ slower than low density CSM in models of typical SNe II \citep{Dessart}. However, lower expansion speeds are characteristic of LLSNe and we are unable to set limits on the density of the CSM from this measurement alone.
 
The H$\alpha$ profile of SN~2018lab becomes complex starting in the 63.9 day spectrum, in the second half of the plateau phase (see Figure \ref{fig:Halpha_evol}). This complex H$\alpha$ profile is not uncommon in LLSNe (see Figure \ref{fig:LLSNe_HaComp}) and has previously been described as the result of the combination of H$\alpha$ and Ba~II $\lambda6497$ \citep{Benetti2001, 05cs, 09N, 08bk, 21aai}. The strength of Ba~II lines in LLSNe is a temperature effect, rather than a relative overabundance. The low temperatures of LLSNe ejecta result in small Ba~III/Ba~II ratios and therefore strong Ba~II lines \citep{Turatto1998}. The presence of exceptionally strong Ba~II lines, particularly Ba~II $\lambda6142$, is a hallmark of the $\sim$80-100 day spectra of LLSNe \citep{Pastorello2004, Spiro2014, Gutierrez2017, Lisakov2018} and is also present in the spectra of SN~2018lab (see Figure \ref{fig:spec}). 

\begin{figure}
    \centering
    \includegraphics[width=\hsize]{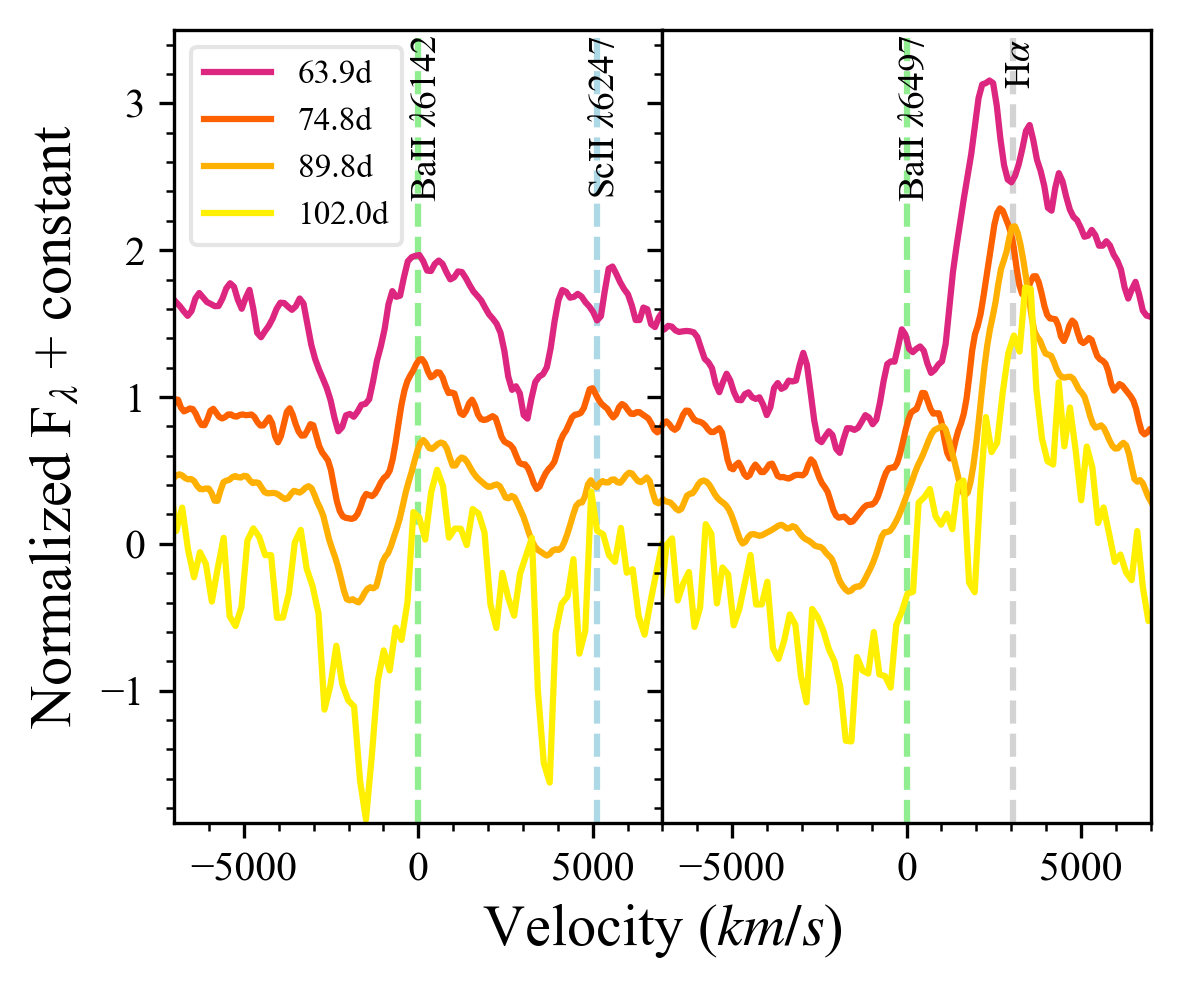}
    \caption{Evolution of the BaII $\lambda6142$ and $\lambda6497$ during the second half of the plateau phase ($\sim50-100$ days). BaII $\lambda6142$ has $v$(BaII)$\approx2000$ km s$^{-1}$. BaII $\lambda6497$ is likely to be evolving similarly to BaII $\lambda6142$, therefore it is very difficult to determine the velocity of H$\alpha$ during these epochs. The relevant spectral lines (dotted) are at rest.} 
    \label{fig:BaII}
\end{figure}

The velocity evolution of Ba~II $\lambda6142$ and $\lambda6497$ is shown in Figure \ref{fig:BaII}. The $v$(Ba~II) of Ba~II $\lambda6142$, the strongest line in the Ba~II multiplet which includes Ba~II $\lambda6497$, is $\sim2000-1500$ km s$^{-1}$ for the $64-102$ day spectra of SN~2018lab. As expected, there is a clear absorption feature centered at $\sim2000$ km s$^{-1}$ in the frame of Ba~II $\lambda6497$ as well. However, the profile of this region makes it difficult to determine the velocity of both Ba~II $\lambda6497$ and H$\alpha$ in the epochs where Ba~II $\lambda6497$ is present. 

Higher signal-to-noise spectra of LLSNe within the crucial second half of the plateau phase are required in order to better understand the structure of the region surrounding H$\alpha$.
Both SYNOW \citep{Pastorello2004, 09N} and CMFGEN \citep{08bk, Lisakov2018} based models of LLSNe spectra fail to adequately replicate the H$\alpha$ profile. Barium (Ba) is an s-process element and is not included in current models. Detailed modeling, which includes Ba~II, of the H$\alpha$ region in LLSNe is needed to facilitate a better understanding of the role of metals on the spectral evolution of LLSNe.  

\begin{figure*}
    \centering
    \includegraphics[width=\hsize]{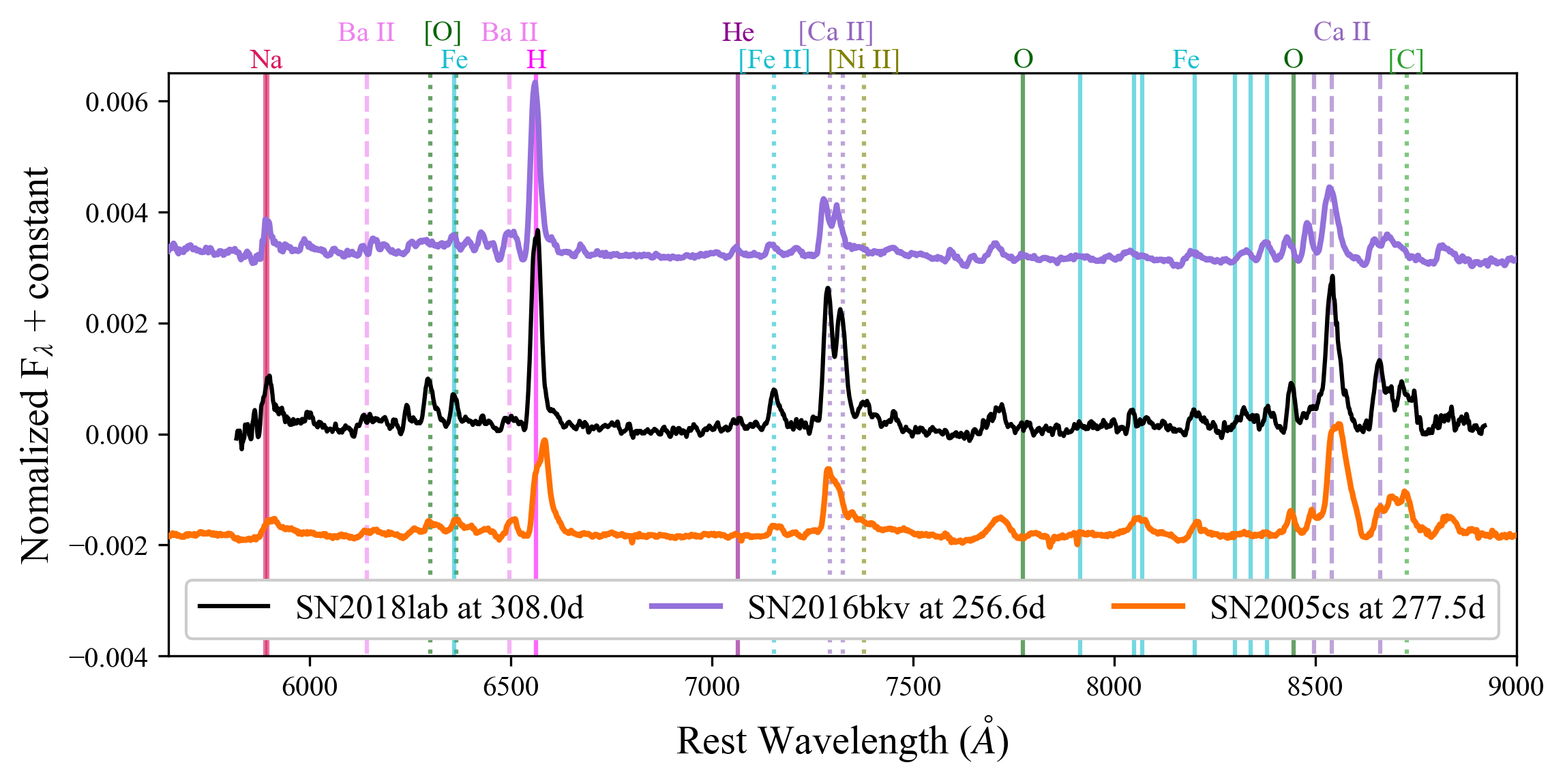}
    \caption{Late-time spectrum of SN~2018lab taken 308 days post explosion compared late time spectra of SN~2005cs \citep{05cs} and SN~2016bkv \citep{16bkv}. The spectrum of SN~2018lab is smoothed using a 10 pixel wide box kernel. All spectra are normalized to the total flux over the wavelength range of the observed SN~2018lab spectrum. While there are many similarities among all three spectra, [C~I] $\lambda8727$ is only present in the nebular spectra of SN~2005cs and SN~2018lab.}
    \label{fig:nebspec}
\end{figure*}

\subsection{Nebular Spectra}\label{sec:nebspec}

Once SN ejecta are predominately transparent to optical light, several clues to the progenitor emerge in the nebular spectra. We obtained a nebular spectrum of SN~2018lab at 308 days post explosion. In Figure \ref{fig:nebspec}, the nebular spectrum of SN~2018lab is compared to similar spectral epochs of SN~2005cs \citep{05cs}, which has a confirmed low mass RSG progenitor, and SN~2016bkv \citep{16bkv}, which has been suggested as a possible ECSN. The SN~2018lab spectrum presented in this figure has been smoothed using a 10 pixel wide box kernel to reduce the appearance of noise. While its nebular spectrum has many of the same features exhibited in both SN~2016bkv and SN~2005cs, SN~2018lab's strong [C~I] $\lambda8727$ feature is only present in the nebular spectrum of SN~2005cs. The importance of this is explained below.

\begin{figure*}
    \centering
    \includegraphics[width=\hsize]{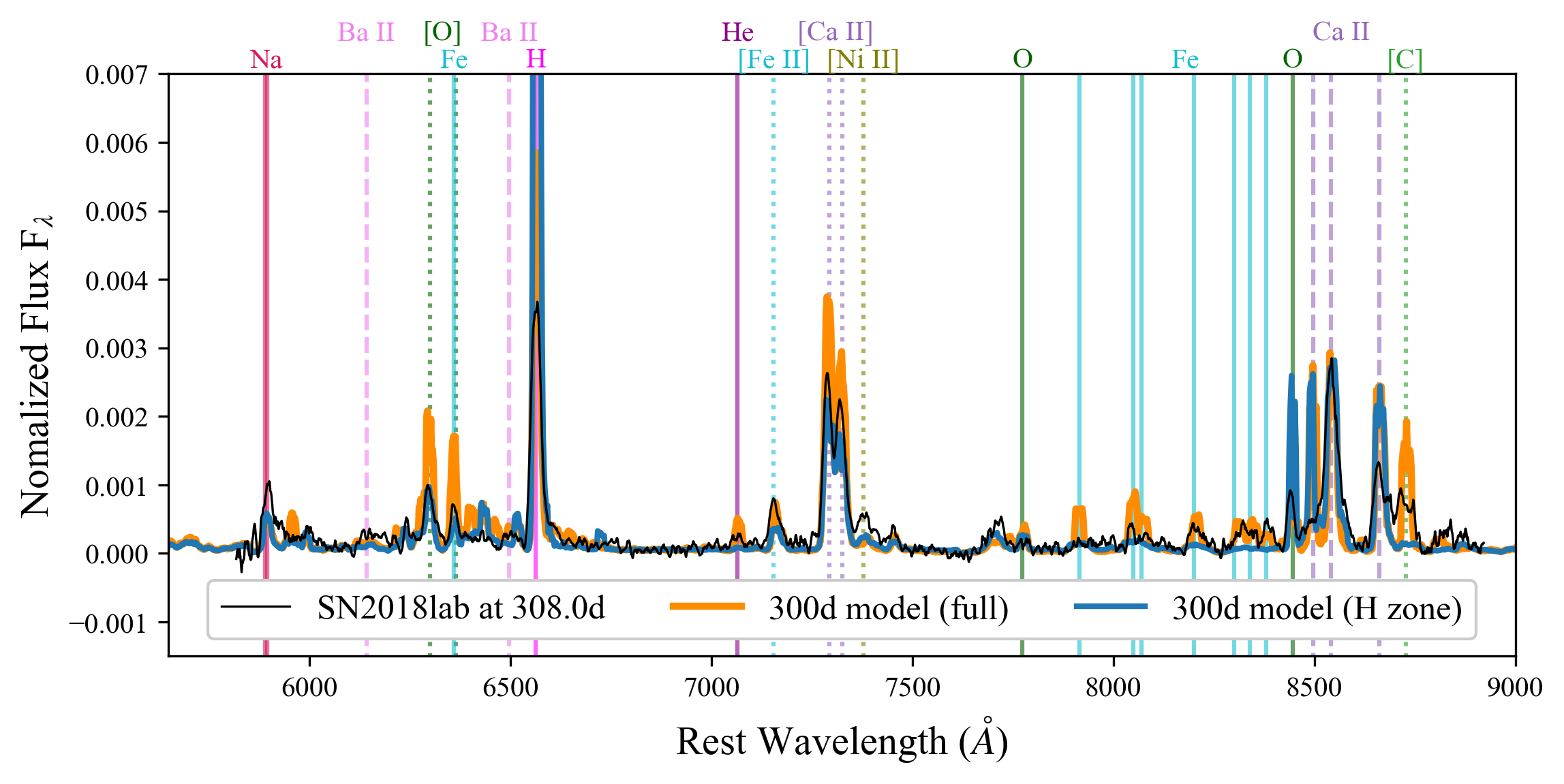}
    \caption{Late-time spectrum of SN~2018lab taken 308 days post explosion compared to the 9 M$_\odot$ \citet{Jerkstrand18} models. Both the models and the spectrum are normalized to the total flux over the wavelength range of the observed spectrum to highlight line ratio differences. The full model, orange, is the expected spectrum for a iron CCSN. The hydrogen-zone model, blue, \citet{Jerkstrand18} should be similar to the nebular spectrum expected of a ECSN. The late-time spectrum of SN~2018lab is similar to that of the iron core collapse model. In particular, [C~I] $\lambda$8727 is present in the spectrum of SN~2018lab and is not included in the ECSNe model.}
    \label{fig:nebmod}
\end{figure*}

In Figure \ref{fig:nebmod}, the nebular spectrum of SN~2018lab is compared to the 300 day nebular spectra models for a 9 M$_\odot$ RSG progenitor presented in \citet{Jerkstrand18}. Since we are unable to determine the nickel mass of SN~2018lab and therefore can not correct for the nickel luminosity at this phase, these models and the spectrum are all normalized to the total flux over the wavelength range of the observed spectrum. The ``pure hydrogen-zone" model presented in \citet{Jerkstrand18} describes the signatures of a progenitor made up of only material from the hydrogen envelope (see their figure 2). While the H-zone model is not a electron-capture model, they expect a ECSN to resemble this model. The full Fe core-collapse model is distinctive from the H-zone model, particularly notable is the lack of He~I $\lambda7065$, Fe~I $\lambda7900-8500$, and [C~I] $\lambda8727$ in the H-zone model. 

SN~2018lab clearly exhibits [C~I] $\lambda8727$ and several Fe~I $\lambda7900-8500$ lines. There is also some evidence of He~I $\lambda7065$. The appearance of these lines, though weaker than indicated by the model, strongly suggests the existence of He and O zones in the progenitor star at the time of collapse. This stellar composition indicates that SN~2018lab is likely to be the result of iron core-collapse in a red supergiant. 
Pre-explosion HST images of the IC~2163/NGC~2207 are unable to offer robust confirmation of this progenitor hypothesis. Given the distance to the host galaxy, the environment surrounding the SN, and the likelihood of a low mass progenitor star, further HST images of the site of SN~2018lab are required to shed light on the progenitor of SN~2018lab and the progenitors of LLSNe in general.

\section{Summary \& Conclusions}\label{sec:summary}

We present comprehensive photometric and spectroscopic observations of SN~2018lab. The early light curve of SN~2018lab is one of the best sampled SNe II to date due to the 30 minute cadence TESS light curve. The TESS light curve combined with extensive photometric and spectroscopic follow up places tight constraints on the early evolution and explosion epoch of SN~2018lab \citep[see also the recent extensive follow-up campaign of the TESS-observed SN~2019esa;][]{19esa}. 

SN~2018lab is among the rare class of LLSNe with observational evidence of short-lived CSM interaction. First, the rising light curve can not be fit with an analytic model of shock cooling \citep{Sapir2017}, indicating that the fast rise is likely the result of excess luminosity due to ejecta-CSM interaction, which is not accounted for in the model. 
Second, the flash spectroscopy in the first couple days following explosion reveals the presence of CSM around the progenitor star. In particular, the broad, ledge-shaped  spectral feature at $\sim$4500--4750\AA~in the +1.9d spectrum of SN~2018lab is analogous to models of ejecta interaction of a RSG with an extended envelope and encompassed by close-in CSM \citep{Dessart}. While we do not explicitly rule out a super-AGB or high mass ($>20 M_\odot$) RSG progenitor, the light curve shape and spectral evolution of SN~2018lab are similar to typical LLSNe, including SN~2005cs which has an identified low mass (10$\pm$3 M$_\odot$) RSG progenitor \citep{Li2006}. Further, the nebular spectrum of SN~2018lab displays many of the features expected to appear in the late-time spectra of iron CCSNe, adding to the likelihood of a RSG progenitor. Given the distance to the host and the nearby H~II region, the pre-explosion HST images of SN~2018lab alone do not set strong enough limits to determine the progenitor of SN~2018lab. Additional post-explosion HST images taken after the SN light has sufficiently faded are required to set the robust constraints on the progenitor of SN~2018lab necessary to test the progenitor pathway suggested in this work.

Currently, there is no indication that the progenitor of SN~2018lab is not a RSG, suggesting that late stage mass loss may be common in LLSNe progenitors regardless if they are RSGs or super-AGBs. Evidence of CSM interaction alone is not enough to determine whether or not a LLSN is the result of electron-capture or core-collapse. Some work has been done to determine the characteristics which distinguish electron-capture from core-collapse processes, including line ratios in nebular spectra and progenitor identification \citep{18zdEC}, but this is still in its early phases and uncertain.
In order to truly understand the progenitor pathways of LLSNe, more spectra and photometry of these objects are urgently needed, not only following explosion but also during the nebular phase. 

SN~2018lab is one of the few LLSNe with observed flash features. The increase in SNe II spectra taken in the hours and days following the explosion has uncovered the diverse morphology in broad early spectral features. Further early observations of SNe II, including the least luminous tails of the SNe II distribution, will shed light on the extent and mechanics of late stage mass loss in RSGs. 

\section*{Acknowlegdements}
We thank Luc Dessart for providing his model spectra.
Time domain research by the University of Arizona team and D.J.S.\ is supported by NASA grant 80NSSC22K0167, NSF grants AST-1821987, 1813466, 1908972, \& 2108032, and by the Heising-Simons Foundation under grant \#2020-1864. 
J.E.A.\ is supported by the international Gemini Observatory, a program of NSF's NOIRLab, which is managed by the Association of Universities for Research in Astronomy (AURA) under a cooperative agreement with the National Science Foundation, on behalf of the Gemini partnership of Argentina, Brazil, Canada, Chile, the Republic of Korea, and the United States of America. Research by Y.D., N.M., and S.V.\ is supported by NSF grants AST-1813176 and AST-2008108.
K.A.B. acknowledges support from the DIRAC Institute in the Department of Astronomy at the University of Washington. The DIRAC Institute is supported through generous gifts from the Charles and Lisa Simonyi Fund for Arts and Sciences, and the Washington Research Foundation.  
The SALT data reported here were taken as part of Rutgers University program 2018-1-MLT-006 (PI: S.~W.~Jha).
This research has made use of the NASA/IPAC Extragalactic Database (NED), which is funded by the National Aeronautics and Space Administration and operated by the California Institute of Technology.
This research has also made use of the Spanish Virtual Observatory \url{https://svo.cab.inta-csic.es}) project funded by MCIN/AEI/10.13039/501100011033/ through grant PID2020\-112949GB\-I00 and the Weizmann Interactive Supernova Data Repository (WISeREP) \citep[\url{https://wiserep.weizmann.ac.il},][]{WISeREP}.
Based in part on data acquired at the Siding Spring Observatory 2.3\,m, we acknowledge the traditional owners of the land on which the SSO stands, the Gamilaraay people, and pay our respects to elders past and present.

\facilities{ADS, CTIO:PROMPT, LBT (MODS), Las Cumbres Observatory (Sinistro, FLOYDS), Keck I (LRIS), Meckering:PROMPT, NED, NTT (EFOSC2), SALT (RSS), Spitzer (IRAC), WISeREP}

\software{  
astropy \citep{astropy:2013, astropy:2018}, corner \citep{corner}, emcee \citep{emcee}, FLOYDS pipeline \citep{FLOYDS}, \texttt{HOTPANTS} \citep{hotpants}, \texttt{lcogtsnpipe} \citep{Valenti2016}, Light Curve Fitting \citep{lightcurvefitting}, MatPLOTLIB \citep{mpl}, NumPy \citep{numpy}, PySALT \citep{SALTpipe}, Scipy \citep{scipy}  
}

\bibliography{sources}
\bibliographystyle{aasjournal}

\end{document}